\documentclass[ba,preprint]{imsart}

\RequirePackage{amsthm,amsmath,amsfonts,amssymb}
\PassOptionsToPackage{hyphens}{url}
\RequirePackage[authoryear]{natbib}
\RequirePackage[colorlinks,citecolor=blue,urlcolor=blue,backref=page,backref=page]{hyperref}
\RequirePackage{graphicx}
\usepackage{longtable,booktabs,array}
\usepackage{float} 
\usepackage{algpseudocode}
\usepackage{algorithm}

\pubyear{2025}
\arxiv{2508.11814}
\volume{TBA}
\issue{TBA}
\firstpage{1}
\lastpage{1}

\startlocaldefs

\newcommand{\extraref}[1]{\ref{#1}}

\newcommand{\myint}[3][]{\int#1 #3 \text{d}#2}

\newcommand{\citeinparen}[1]{\citeauthor{#1}, \citeyear{#1}}

\newcommand{\historyinfoblue}{The horizontal blue line marks rejecting perfect calibration with 5\% false positive rate.}
\newcommand{\historyinfo}{\historyinfoblue\ The vertical
orange line marks when the check first attains
80\% power. }

\newcommand{\calibrationplotinfo}{horizontal axis is the predicted probability for
model 1, vertical axis is the conditional event probability (CEP) that model 1 generated the data. Overall trend (red) and confidence region (blue) as in \cite{dimitraidis2021_reliability}. A histogram of observed probabilities is shown at the bottom. 
}

\newcommand{\figurescale}{1}

\providecommand{\tightlist}{%
  \setlength{\itemsep}{0pt}\setlength{\parskip}{0pt}}

\theoremstyle{plain}

\newtheorem{theorem}{Theorem}

\theoremstyle{definition}
\newtheorem{definition}{Definition}

\newtheorem{remark}[theorem]{Remark}
\theoremstyle{remark}

\endlocaldefs

\begin{document}

\begin{frontmatter}
\title{Simulation-based validation of Bayes factor computation}
\runtitle{Simulation-based validation of Bayes factors}

\begin{aug}
\author[mm]{\fnms{Martin} \snm{Modr\'ak}\ead[label=e1]{modrak.mar@gmail.com}}\orcid{0000-0002-8886-7797},
\author{\fnms{Sebastian} \snm{Stroppel}\thanksref{ss}}\orcid{0009-0002-5239-2037},
\and
\author{\fnms{Paul} \snm{B\"{u}rkner}\thanksref{pb}}\orcid{0000-0001-5765-8995}

\runauthor{M. Modr\'ak et al.}

\address[mm]{Department of Bioinformatics, Charles University, Prague, Czech Repbulic\printead[presep={,\ }]{e1}}
\address[ss]{Department of Mathematical Stochastics, University of Freiburg, Germany}
\address[pb]{Department of Statistics, Technical University of Dortmund, Germany}

\end{aug}



\begin{abstract}

We propose and evaluate two methods that validate the computation of Bayes factors: one based on an improved variant of simulation-based calibration checking
(SBC) and one based on calibration metrics for binary predictions. We show that in theory, binary prediction calibration is equivalent to a special case of SBC, but with limited resources, binary prediction calibration is typically more sensitive to the problems we investigated. With well-designed test quantities, SBC can however detect all possible problems in computation, including some that cannot be uncovered by binary prediction calibration. 

Previous work on Bayes factor validation includes checks based on the data-averaged posterior and the Good check method. We demonstrate that both checks miss many problems in Bayes
factor computation detectable with SBC and binary prediction calibration. Moreover, we find that the Good check as originally described fails to control its error rates. Our proposed checks also typically use simulation
results more efficiently than data-averaged posterior checks. 
Finally, we show that a special approach based on posterior SBC is
necessary when checking Bayes factor computation under improper priors and we validate several models with such priors. 

We recommend that novel methods for Bayes factor computation be validated with SBC, binary prediction calibration and data-averaged posterior with at least several hundred simulations. For all the models we tested, the \texttt{bridgesampling} and \texttt{BayesFactor} R packages satisfy all available checks and thus are likely safe to use in standard scenarios.

\end{abstract}



\end{frontmatter}

\newcommand{\refextrafigure}[1]{Appendix E, Figure~\extraref{#1}}
\newcommand{\refextratable}[1]{Appendix E, Table~\extraref{#1}}

\section{Introduction}\label{introduction}

Bayes factors (BFs), i.e.~the ratio of marginal likelihoods are a
commonly used quantity for model selection and hypothesis testing in the
Bayesian paradigm. BFs have a number of theoretically appealing properties: 
they arise naturally when the model is treated as uncertain in a Bayesian paradigm, they can be understood as measuring the relative weight of evidence, they ensure coherent assignment of probabilities to events and allow for sequential updating (see e.g., \citeinparen{Wagenmakers2018_bf_advantages} for a broader discussion).
In addition to these benefits mostly requiring an $\mathcal{M}$-closed setting (that the set of possible models of the data is known, see \citeinparen{key1999_model_choice}),
BFs have also faced other criticisms including: 
sensitivity to prior choice \citep{LIU200BFSensitivity}, 
the Jeffreys-Lindley paradox \citep{Wagenmakers2023_jf_paradox}, providing counter-intuitive 
results under the $\mathcal{M}$-open setting (when none of the compared models is correct, see \citeinparen{yao_stacking_2018}),
or that rules of thumb for interpreting the weight of evidence based on Bayes factors are
not well justified as decision rules \citep{schad_workflow_2023}.

In this paper, we set the discussion about the usefulness of Bayes
factors aside, and simply assume we have a scenario where Bayes factors
are a useful quantity of interest. There still remains a practical
concern: how to compute Bayes factors 
as this typically involves 
high-dimensional integrals that 
need to be numerically approximated. 

A simple approach, applicable only for nested models where one
parameter is fixed to some constant, is the Savage-Dickey method \citep{dickey1970}. For
specific families of models, the integration problem can be simplified
to a numerically tractable low-dimensional integral, 
as implemented in the \texttt{BayesFactor} package \citep{bayesfactor_package}. A
flexible method is bridge sampling \citep{meng_warp_2002}, 
implemented in the \texttt{bridgesampling} package
\citep{bridgesampling_package}, which can work with any
set of samples from model posterior to provide an approximation to the
marginal likelihood (and thus Bayes factors). Further methods are discussed in the reviews by \cite{raftery1995bayesfactors}, \cite{han2001bfmcmc} and \cite{llorente2023bfreview}.
Recent work includes the learned harmonic mean estimator \citep{mcewen2023machinelearningassistedbayesian} and methods for calculating marginal likelihoods with simulation-based inference \cite[e.g., ][]{spuriomancini2023bfsbi,radev2023bfsbi,Srinivasan2024bfsbi,polanska2025bfsbi}.
However, the validity of
those approximations for specific model selection scenarios is only rarely tested.

Our main contribution is to employ an improved variant of
simulation-based calibration checking (SBC, \citeauthor{modrak2025} \citeyear{modrak2025}) and the
accompanying theoretical progress for validating Bayes factor
computation. On the theoretical side, we show an equivalence between SBC for Bayes
factors and binary prediction calibration and that the approach is valid also when focusing on a subset of the full data space. 
On the practical side, we describe the specifics of using SBC as well as posterior SBC  \citep{sailynoja2025posteriorsbc} for Bayes factor validation with real-world computational tools and compare the performance of SBC with other checks.
Using several toy models, we
demonstrate how problems in Bayes factor computation not detectable by
previously reported methods can easily be detected by SBC. Finally, we
turn to real-world problems and show examples of full validation
workflow for bridge sampling over Stan models and compare efficiency of
various checks in discovering errors in normalization constants.
We also demonstrate that using posterior SBC
is necessary when checking Bayes
factor computation under improper priors for some model parameters. This
in turn allows us to validate the computation for several models
available in the \texttt{BayesFactor} package. Additionally, we show an example 
comparing multiple models at once and detecting incorrect computations for the Savage-Dickey density ratio.

Previously, Schad and co-authors have developed a workflow for
validating Bayes factor computation inspired by SBC and applied it to
several problems \citep{schad_workflow_2023,schad_aggregation_2024}. 
From the theoretical side, we note that the Schad et
al.~approach differs importantly from SBC both as originally described
\citep{talts_sbc} as well as our improved variant \citep{modrak2025}. In fact, the 
original Schad et al.~approach is better understood as a check based on the data-averaged
posterior. As we show in this paper, data-averaged posterior checking typically uses
simulation results inefficiently and is insensitive to large classes of
potential problems.

While this paper was in preparation, we collaborated with Schad \& Vasishth on combining data-averaged
posterior checking and binary prediction calibration to validate Bayes factor for
several families of models popular in cognitive science \citep{schad2025bf}, employing results of an earlier version of this paper.

In the remainder of the paper, we first provide an overview of known results (Section \ref{theoretical-background}), 
then we discuss our new theoretical contributions to the validation of Bayes factor computations 
(Section \ref{theoretical-contributions}). After describing our evaluation methods
(Section \ref{evaluation-methods}),
we present simulation results on toy examples (Section \ref{toy-examples}) 
and then move to more realistic use cases (Section \ref{realistic-examples}).
We conclude the paper with recommendations for checking Bayes factor 
computation (Section \ref{discussion}). 
Full
code to reproduce all simulation results in this paper can be found at
\url{https://github.com/martinmodrak/SBCBayesFactors/}.

\section{Theoretical background and related work}\label{theoretical-background}

Throughout this paper we assume there is a shared data space
\(Y\) and a set of models \(\mathcal{M}_i\) for
\(i \in \{0, \dots, K-1\}\), each model is associated with parameter
space \(\Theta_i\). For \(y \in Y, \theta^i \in \Theta_i\) each model
implies the following joint, marginal, and posterior densities (we use ``density'' to 
refer to both density and mass functions as well as their combinations):
\begin{align}
    \pi^i_\text{joint}(y, \theta^i) &= \pi^i_\text{obs}(y \mid  \theta^i) \pi^i_\text{prior}(\theta^i)  &  \pi^i_\text{marg}\left(y \right) &= \myint[_\Theta]{\theta^i}{\pi^i_{\text{obs}}(y \mid  \theta^i) \pi^i_\text{prior}(\theta^i)} \notag \\
    \pi^i_\text{post}(\theta^i \mid  y) &= \frac{\pi^i_\text{obs}(y \mid  \theta^i) \pi^i_\text{prior}(\theta^i)}{\pi^i_\text{marg}\left(y \right)} & & .
\end{align}

When there is no risk of confusion we may omit the model index. We will
use
\(\pi^i_\text{marg}, \pi^i_\text{prior}, \pi^i_\text{obs}(\theta), \pi^i_\text{post}(y)\)
to denote the distributions themselves. We
will use \(\phi_\text{post}\) and \(\phi_\text{marg}\) to denote
candidate densities that are not possibly incorrect.

\subsection{Bayesian model averaging, Bayes Factors}\label{bayesian-model-averaging}

Bayes factors are closely related to Bayesian model averaging (BMA). In
BMA, we assume a supermodel for the data with observation space \(Y\)
and parameter space
\(\Xi = \{ (0, \theta_0 ) \mid \theta_0 \in \Theta_0\} \cup \dots \cup  \{ (K-1, \theta_{K-1} ) \mid \theta_{K-1} \in \Theta_{K-1}\}\)
of the form:
\begin{align}
i &\sim \text{Categorical}(\mbox{Pr}(\mathcal{M}_0), \dots, \mbox{Pr}(\mathcal{M}_{K - 1}) ) & & \notag \\
\theta^i \mid i &\sim \pi^i_\text{prior} & 
y \mid i, \theta^i &\sim \pi_\text{obs}^i(\theta^i) &
\end{align}
Given the prior distribution over models, the posterior distribution
\(\pi_\text{BMA}(i \mid y)\) of the model index \(i\) is then fully
determined by the marginal density (\(\pi_\text{marg}\), also known as
marginal likelihood) and vice versa:
\begin{equation}
\mbox{Pr}(\mathcal{M}_i \mid y) = \pi_\text{BMA}(i \mid y) = \frac{\pi^i_\text{marg}(y)\mbox{Pr}(\mathcal{M}_i)}{\sum_{j=0}^{K - 1} \pi^j_\text{marg}(y)\mbox{Pr}(\mathcal{M}_j)}
\end{equation}
And finally the posterior distribution of any quantity of interest is
obtained by weighing the predictions of the individual models, i.e.~for
any quantity \(\Delta\):
\begin{equation}
\pi_\text{BMA}(\Delta \mid y) = \sum_{i=0}^{K - 1} \pi^i_\text{post}(\Delta \mid y) \pi_\text{BMA}(i \mid y)
\label{eq:bma_post}
\end{equation}
This weighted prediction is then typically the output of BMA. 

Given \(y \in Y\), the Bayes factor (BF) of model \(i\) against model
\(j\) can be understood as both the ratio of marginal densities of
the data and the ratio of posterior probabilities of the models relative
to the ratio of their prior probabilities:

\begin{equation}    
BF_{i,j} = \frac{ \pi^i_\text{marg}\left(y \right) }{\pi^j_\text{marg}\left(y \right)} = 
\frac{\mbox{Pr}(\mathcal{M}_i \mid  y)}{\mbox{Pr}(\mathcal{M}_j \mid y)} \frac{\mbox{Pr}(\mathcal{M}_j)}{\mbox{Pr}(\mathcal{M}_i)}
\end{equation}

We further
note that
\begin{equation}
\frac{\pi_\text{BMA}(i \mid y)}{\pi_\text{BMA}(j \mid y)} = \frac{\mbox{Pr}(\mathcal{M}_i \mid y)}{\mbox{Pr}(\mathcal{M}_j \mid y)} = 
\frac{\mbox{Pr}(y\mid \mathcal{M}_i) \mbox{Pr}(\mathcal{M}_i)}{\mbox{Pr}(y \mid \mathcal{M}_j) \mbox{Pr}(\mathcal{M}_j)}   =
\frac{\pi^i_\text{marg}(y) \mbox{Pr}(\mathcal{M}_i)}{\pi^j_\text{marg}(y) \mbox{Pr}(\mathcal{M}_j)}   =
BF_{i,j}\frac{\mbox{Pr}(\mathcal{M}_i)}{\mbox{Pr}(\mathcal{M}_j)}
\end{equation}
Due to this tight connection between Bayes factors, marginal likelihood,
and BMA, computing one immediately lets us derive the others. And so a
check of correctness for either is also a check of correctness for the
others.

\subsection{Bayesian model consistency via
simulations}\label{bayesian-model-consistency-via-simulations}

We can use simulations to validate a Bayesian statistical model --- in our case the implied BMA supermodel. Two classes of
approaches to checking Bayesian computation in general are found in the
literature: simulation-based calibration checking (SBC) and data-averaged posterior checking.
Implicitly, all the theory below holds only
for infinitely precise computation and on average over infinitely many
simulations.

\subsubsection{Simulation-based calibration
checking}\label{simulation-based-calibration-checking}

SBC relies on the joint distribution of the prior and posterior samples
\begin{equation}
\pi_\text{SBC}(y, \theta, \tilde\theta) = \pi_\text{prior}(\tilde\theta) \pi_\text{obs}(y \mid \tilde\theta) \pi_\text{post}(\theta \mid y),
\end{equation}
from which it follows that, conditional on \(y\), \(\tilde\theta\) and
\(\theta\) are exchangeable:
\begin{equation}
\forall y \in Y, \theta, \tilde\theta \in \Theta: \pi_\text{SBC}(y, \theta, \tilde\theta) = \pi_\text{SBC}(y, \tilde\theta, \theta).
\end{equation}
In other words, if we simulate data, we can treat the prior draw $\tilde\theta$ that generated $y$ as a
single draw from the correct posterior $\pi_\text{post}(y)$. Intuitively, $\tilde\theta$ must (on average) be compatible with $y$, i.e. be representative of the posterior. To phrase it yet differently, if the prior draw $\tilde{\theta}$ that generated the data $y$ is mixed among draws from $\pi_\text{post}(y)$ we could do no better than chance in figuring out which draw originated from the prior and which from the posterior.

The exchangeability implied by the equation above cannot be checked
directly, so we make two changes: 1) check on average over \(Y\) and 2)
use a test quantity
\(f: \Theta \times Y \to \mathbb{R} \cup \{-\infty,\infty\}\). When our
posterior is represented by samples (e.g.~via Markov-chain Monte Carlo),
then for each simulation we compute the ranks of the prior draw within
the samples and note that those ranks need to follow a discrete uniform
distribution --- we call this process \emph{sample SBC}. When the
cumulative distribution function (CDF) of the posterior is available 
(e.g.~from a variational approximation,
or a direct estimate of posterior model probability), we can take the CDF
value at the prior draw for each simulation and those values need to follow a
continuous uniform distribution --- we call this \emph{continuous SBC}.
Since \(f\) is used only for the ordering it implies, any
totally-ordered set will do as the target of a test quantity. This is
identical to checking that for all \(0 < p < 1\), all posterior credible
intervals of width \(p\) contain the simulated value \(p\) of the time.
When ties are possible, we need to pick the rank uniformly at random within
the range of the ties. 

More formally, $M$-sample SBC proceeds as follows \citep{modrak2025}:
\begin{equation}
\begin{aligned}
  {\tilde\theta} &\sim \pi_\text{prior}    \\
  {y} \mid \tilde\theta &\sim \pi_\text{obs}({\tilde\theta})    \\
  \theta_1, \dots \theta_M \mid y &\sim \phi_\text{post}({y}),
  \label{eq:sbc_setup}
\end{aligned}
\end{equation}
We then define:
\begin{equation}
\begin{aligned}
   N_{\mathtt{less}} &:= \sum_{m=1}^M \mathbb{I} \left[f(\theta_m, y) < f(\tilde \theta, y) \right] \\
   N_{\mathtt{equals}} &:= \sum_{m=1}^M \mathbb{I} \left[f(\theta_m, y) = f(\tilde \theta, y) \right] \\
   K \mid N_{\mathtt{equals}} &\sim \mathrm{uniform}(0,  N_{\mathtt{equals}})\\
   N_\mathtt{total} &:= N_{\mathtt{less}} + K,
\end{aligned}  
\end{equation}

Then, if the candidate posterior $\phi$ (e.g. implemented in a probabilistic program) and the generator $\pi$ implement the same probabilistic model, we have
\begin{equation}
    N_{\mathtt{total}} \sim \mathrm{uniform}(0, M)   
    \label{eq:sbc_sample}
\end{equation}

and SBC relies on checking discrete uniformity of $N_\mathtt{total}$ empirically. 
Continuous SBC then arises 
in the limit as $M \to \infty$, where $\frac{N_\mathtt{total}}{M}$ now follows a continuous uniform distribution on $[0,1]$.
\citet{modrak2025}, Theorems 2 and 3 show that checking continuous SBC is equivalent to checking $M$-sample SBC for all $M$.

An important property of SBC is that, by allowing test quantity \(f\) to
depend on \(y\), we can in principle detect any mismatch between
\(\phi_\text{post}\) and \(\pi_\text{post}\) --- we only need a suitable
test quantity that is sensitive to the problem at hand (\citeauthor{modrak2025}, Theorem 6). \citeauthor{modrak2025}~provide theoretical and empirical evidence that the likelihood
\(f(\theta, y) = \pi_\text{obs}(y \mid \theta)\) is often highly
sensitive in that regard. 

\citet{yao_domke_classifier_sbc_2023}
show that, instead of choosing specific test quantities,
we can train a classifier to distinguish prior and posterior
draws and this allows sensitivity to any type of problem without
explicitly constructing a test quantity. Unfortunately, this comes at the
cost of requiring a potentially very large number of simulations. There
are also other checks following similar ideas, which are reviewed in
\citet{modrak2025} and not discussed here further due to their close resemblance to SBC.

\subsubsection{Posterior SBC}\label{posterior-sbc}

Good choice of priors is important for practical success with SBC. There
are at least three problems one can encounter in this regard: 
\begin{enumerate}
   \tightlist
    \item 
Some
statistical packages for Bayes factors use improper priors, which we cannot simulate from.

 \item Many priors are unrealistic --- notably using independent priors for
all parameters often results in large prior probability on unrealistic
datasets (for details of the problem and some solutions see \citet{hem_makemyprior_2022,aguilar2025generalized,fazio2025primedpriorssimulationbasedvalidation}). 
This both reduces relevance of our simulations for real
datasets and may result in convergence problems for simulated data even
when the model works well enough for real datasets. 
\item If we aim to
verify Bayes factor computation for a specific dataset, even a very good
prior will result in most simulated datasets being very different from the
dataset we are interested in. This makes it less likely to detect potential
problems that manifest only in the particular region of the data space.
\end{enumerate}
All three problems can be overcome with \emph{posterior SBC} \citep{sailynoja2025posteriorsbc}.
The overall idea of posterior SBC is to fit the model to a well-chosen
dataset(s) and treat the resulting posterior(s) as the prior for
simulations. More formally, if we assume the data space partitions as
\(Y = Y_1 \times Y_2\), then given a fixed dataset \(y_1 \in Y_1\) we
can construct a new model \(\bar\pi\) such that:
\begin{equation}
\begin{aligned}
\bar\pi_\text{prior}(\theta) &=  \pi_\text{post}(\theta \mid y_1) \\
\bar\pi_\text{post}(\theta \mid y_2) &= \pi_\text{post}(\theta \mid (y_1, y_2)) \\
\end{aligned}
\end{equation}
We can then simulate from \(\bar\pi_\text{prior}\) by fitting \(\pi\) to
\(y_1\) and taking draws from \(\pi_\text{post}(\theta\mid y_1)\). We
may then use this parameter draw to simulate \(y_2\) from
\(\pi_\text{obs}\) and fit \(\pi\) to the whole dataset \((y_1,y_2)\).
This setup can be generalized further by replacing the single fixed
\(y_1\) with a distribution \(g\) over possible \(y_1\), so we have
\(\bar\pi_\text{prior}(\theta) = \myint[_{Y_1}]{y_1}{g(y_1) \,  \pi_\text{post}(\theta \mid y_1)} \).

To check computation for models with improper priors, we need to choose
\(y_1\) such that \(\pi_\text{post}(\theta \mid y_1)\) is proper and
amenable to sampling. To provide a better prior, we take the
distribution \(g\) as representing prior beliefs about the data (which
tend to be easier to elicit than priors on parameters \citep{hartmann_prior_eliciation}). To
target an SBC check to a specific dataset, we may take a suitable
portion of the dataset as fixed \(y_1\) or even take \(g\) to be a
suitable subsampling distribution of the dataset.

The core idea of posterior SBC (use data to improve prior) is similar to the motivation behind the intrinsic Bayes factor \citep{berger_pericchi_intrinsinc_bf_1996}. There are however also large conceptual differences: 1) For intrinsic BF computation, we use real observed data to improve prior, whereas for posterior SBC, the data need not be real and can instead represent domain knowledge about the expected data values. 2) For intrinsic BF, using data as part of the prior aims to provide a better model comparison result, but in posterior SBC it is just a tool for testing a model, which we then intend to use with unchanged (improper or otherwise imperfect) prior and 3) for intrinsic BF the dependency of results on the choice of specific data to incorporate into the prior is considered undesirable and the authors suggest several ways to avoid it, while for posterior SBC the specific choice would typically be of lower importance --- it just needs to move the simulations roughly into the correct neighborhood, either close to a real dataset of interest or matching prior assumptions about that data.

\subsubsection{Data-averaged posterior checking}\label{data-averaged-posterior}

Another set of approaches to model consistency is based on the fact that the data-averaged
posterior should equal the prior:
\begin{equation}
    \forall \theta: \pi_\text{prior}(\theta) = \myint[_Y]{y} {\phi_\text{post}(\theta \mid y )\myint[_\Theta]{\tilde\theta}{\pi_\text{obs}(y \mid  \tilde\theta) \pi_\text{prior}(\tilde \theta)}}
\end{equation}
That this property holds when $\phi = \pi_\text{post}$ is a direct consequence of the law of iterated expectation $E_Y(E(X | Y = y)) = E(X)$ for any random variable $X$. The prior has to have all expectations identical to the data-averaged posterior and thus has to be an identical distribution.

This equality can once again can be tested by fitting the model to datasets
simulated from the prior and observational model. This method traces
back to \citet{geweke_getting_2004}. 
Typically, we do not test the full joint distribution but suitable
univariate projections \(f: \Theta \to \mathbb{R}\) as there are
practical tests for (in)equality of arbitrary univariate distributions.
Alternatively, we can compare the moments of the full prior to the
moments of the full data-averaged posterior \citep{yu_assessment_2021}.

\citet{schad_workflow_2023} propose a method to test Bayes factor computation which they call SBC.
However, it in fact consists of data-averaged
posterior checking for the model index in the BMA supermodel (i.e.~with
\(f: \Xi_\text{BMA} \to \{0, \ldots K - 1\}\) and
\(\forall (i, \theta_i) \in \Xi: f((i, \theta_i)) = i\)),
although they always compare two models, making the
response actually binary. The situation is simplified by the fact that
both the prior and the data-averaged posterior are Bernoulli distributions and can
thus be fully described by their means.  \citet{schad_workflow_2023}~propose to use Bayesian t-test for the mean comparison.

Data-averaged posterior checking is repeatedly conflated with SBC in the
literature on model validation, although they are fundamentally different
methods. In full generality, data-averaged posterior checking and SBC are not
comparable --- there are test quantities and candidate posteriors that
will pass one but fail the other (see \citeinparen{modrak2025} for examples).

\subsection{Good check}\label{good-check}

The
\emph{Good check} \citep{Sekulovski2024goodcheck} was recently proposed as another method to test Bayes factor
computation. It notes that the expected Bayes factor in favor of the
wrong hypothesis is 1, i.e.:
\begin{equation}
\mathbb{E}(BF_{0,1} \mid  \mathcal{M}_1) = 1,
\end{equation}
which is a special case of a more general identity for moments of Bayes
factors:
\begin{equation}    
\forall k \in \mathbb{N}:\mathbb{E}(BF^{k + 1}_{0,1} \mid  \mathcal{M}_1) = \mathbb{E}(BF^k_{0,1} \mid  \mathcal{M}_0).
\end{equation}
\citet{Sekulovski2024goodcheck}~recommend empirical testing of the identity for
\(k = 0\) and \(k = 1\) via simulations, but offer only an informal visual test of the identities, using tens of thousands of simulations. 
They also note that the Bayes factor is ``likely to have a (roughly) log-normal distribution'' and convergence to the expected value may be slow in some cases. They note that among the two possible directions of the Good check,
simulating from the more general of the two models leads to better convergence.

We were made aware of a currently unpublished modification of the check that should provide improved behavior (personal communication with E.J. Wagenmakers). We expect to evaluate this extension once it is made public.

\section{Theoretical contributions}
\label{theoretical-contributions}

We offer binary prediction calibration as a previously unused method to validate Bayes factor computation and show its close connection to SBC. Beyond this connection our contributions to SBC are 1) working out the specifics of using posterior SBC for Bayes factor validation and 2) highlighting how it can be used when focusing on a subset of the full data space. We also discuss new insights into the data-averaged posterior check and the Good check.

\subsection{Bayes factors and binary prediction
calibration}\label{bayes-factors-and-binary-prediction-calibration}

There is a straightforward, but as of yet unreported way to check the correctness of posterior
model probabilities with respect to the BMA supermodel --- whenever
\(\pi_\text{BMA}(i \mid y) = p\), the true model should be
\(\mathcal{M}_i\) in \(p\) of the cases. We say that, in this case,
the predictions are \textit{calibrated}.

Formally, given a candidate (possibly incorrect) density
\(\phi_\text{BMA}(i \mid y)\), denote the posterior model probability 
$D_k(y) = \mbox{Pr}_{\phi_\text{BMA}}[i = k \mid y]$. This quantity can be treated as a random variable with respect to the model $\pi$. \(\phi_\text{BMA}\) satisfies the
\emph{binary prediction calibration check} w.r.t. the model index if:
\begin{equation}
\forall p \in [0,1], k \in \{0, \dots, K - 1\} : \mbox{Pr}_\pi[i = k\mid D_k = p] = p,    
\end{equation}
wherever the left-hand side is well-defined (i.e.~there exists data \(y\)
within the support of the BMA supermodel for which \(\phi\) predicts
posterior probability \(p\) for model \(k\)).

This property can be empirically checked via simulations: For
each simulation \(s \in \{1, \dots, S\}\), we first choose the model
according to model prior:
\begin{equation}
  i_s \sim \text{Categorical}(\mbox{Pr}(\mathcal{M}_0), \dots, \mbox{Pr}(\mathcal{M}_{K - 1}) ),  
\end{equation}
where the categorical distribution is analogous to a multinomial distribution with one trial. We then draw from the prior and observational distribution to obtain
\(y_s \sim \pi^{i_s}_\text{marg}\) and compute
\(\phi_\text{BMA}(j \mid y_s)\) for \(j \in \{0, \dots, K - 1\}\). The $i_s$ then serve as true values to check the calibration of predictions made by $\phi_\text{BMA}(j \mid y_s)$ using any suitable method.

Especially for binary case (\(K = 2\)) there is a number of
available methods for checking binary prediction calibration (see e.g., \citeauthor{dimitriadis2023_calibration} \citeyear{dimitriadis2023_calibration}%
). The \(K > 2\) case can then be handled by looking at the
binary calibration of the top (highest probability) prediction or by
looking at a set of \(K\) binary calibrations of the probability of one
model against all other models.

Binary prediction calibration of the model index will however be satisfied even if
\(\phi_\text{BMA} \neq \pi_\text{BMA}\) --- most notably, the check is
satisfied if we ignore the data
\(\phi(i \mid y) = \mbox{Pr}(\mathcal{M}_i)\) or do not use the data
completely (see Remark 2 in Appendix A).

\subsection{Use of SBC for the BMA supermodel}

Applying SBC to the BMA supermodel provides a direct way to check Bayes factor computation and is a novel 
contribution of this paper. While the BMA supermodel is rarely fit directly, it can be easily reconstructed from the  candidate posterior distributions for the individual models and the posterior model probabilities, specifically to take draws from the BMA posterior we first draw the model index and then pick a draw from the posterior of the chosen submodel:
\begin{equation}
\begin{aligned}
    i_\text{post} &\sim \text{Categorical}(\phi_\text{BMA}(0 \mid y), \dots, \phi_\text{BMA}(K \mid y)) \\
    \theta^i_\text{post} \mid i_\text{post} &\sim \phi^{i_\text{post}}(y).  \label{eq:bma_draws}
\end{aligned}
\end{equation}
Where $\phi^{i}(y)$ is the candidate posterior distribution of the $i$-th submodel given $y$. Note that \eqref{eq:bma_draws} is just a restatement of the claim about densities in \eqref{eq:bma_post}. The SBC process has large overlap with the simulation setup for checking binary calibration --- we draw the model index from the model prior, draw parameters from the prior of the selected model, draw data from the relevant observational distribution, fit the models and compute posterior model probabilities. The difference is only in the post-processing of the results. 
A pseudocode representation of SBC for Bayes factors as implemented in the \texttt{SBC} package can be found in Algorithm~\ref{sbcalg}. We show the two models case ($K = 2$) for simplicity, extension to more models is possible by 1) drawing the model index from a categorical distribution for both prior draw and posterior computation and 2) replacing the explicit handling of two models with iteration over multiple models. Some more details on the possible choices of test quantities ($f_1, \dots, f_J$) are given in Appendix B.

\begin{algorithm}[bt!]
    \caption{SBC for Bayes factor/BMA with two models}
    \label{sbcalg}
    \begin{algorithmic}
        \Function{RankFromDraws}{$x$ : simulated value, $d$ : array of posterior draws}
           \State $l \gets $ no. of elements of $d$ less than $x$
           \State $e \gets $ no. of elements of $d$ equal to $x$
           \State \Return integer uniformly at random between $l$ and $l + e$
        \EndFunction
        \State
        
        \Function{SingleSimulation}{$M$: number of draws taken, $f$: a vector of $J$ test quantities}
           \State draw $i \sim \text{Bernoulli}(\mbox{Pr}(\mathcal{M}_{1}))$
           \State draw $\tilde\theta \sim \pi^i_\text{prior}$
           \State draw $y \sim \pi_\text{obs}^i(\tilde\theta)$
           \If{$y$ is not acceptable}
              \State go to start of the function \Comment{Rejection sampling of datasets} 
           \EndIf
           \State draw $\theta^{0}_1, \dots, \theta^{0}_M \sim \phi^0(\theta \mid y)$ 
           \State draw $\theta^{1}_1, \dots, \theta^{1}_M \sim \phi^1(\theta \mid y)$ 
           \State $b \gets BF_{01}$ according to the tested algorithm           
           \State $z\gets \frac{1}{b}\frac{\mbox{Pr}(\mathcal{M}_1)}{\mbox{Pr}(\mathcal{M}_0)}$, $p \gets \frac{z}{z + 1}$
           \State draw $m_1, \dots, m_M \sim \text{Bernoulli}(p)$
           \State $r^i \gets $ \Call{RankFromDraws}{$i$, $m$} \Comment{rank for the binary model index}
           \For{j := 1, dots, J}
              \State $f_\text{vec} \gets f_j((m_1, \theta^{m_1}_1), y), \dots, f_j((m_M, \theta^{m_M}_M), y)$
              \State $r^f_j \gets $ \Call{RankFromDraws}{$f_j((i, \tilde\theta), y)$, $f_\text{vec}$} \Comment{rank for other test quantities}
           \EndFor
           \State \Return $(i, p, r^i, r^f)$
        \EndFunction
        \State
        
        \Procedure{SBC\_BF}{$M$: number of draws taken, $f$: a vector of $J$ test quantities}
           \For{s := 1, \dots, S}
              \State $(i_s, p_s, r^i_s, r^f_{(s,)}) \gets$ \Call{SingleSimulation}{M, f}
           \EndFor
           \State test $r^i$ uniformly distributed over $1, \dots, M$
           \Comment{SBC for model index}
           \For{$j := 1, \dots, J$}
              \State test $r^f_{(,j)}$ uniformly distributed over $1, \dots, M$
              \Comment{SBC for other test quantities}
           \EndFor
           \State test $\mathbb{E}[p] = \mbox{Pr}(\mathcal{M}_{1})$  \Comment{Data-averaged posterior}
           \State test $\mathbb{E}[i \mid p] = p$  \Comment{Binary prediction calibration}
        \EndProcedure
    \end{algorithmic}
\end{algorithm}

Since SBC can
in principle detect any problem in any Bayesian model, it can detect any problem in Bayes factor computation as long as suitable test quantities are employed. 
We further note that when \(K = 2\), then SBC for the model index in the BMA supermodel is in fact the same as binary
prediction calibration for the model index, i.e.~considering a candidate posterior for the
BMA supermodel, \(\phi_\text{BMA}\) passes continuous SBC w.r.t.
\(f: \Xi \to \{0, 1\}, f((i, \theta_i), y) = i\) if and only if
\(\phi_\text{BMA}\) passes the binary prediction calibration check w.r.t. the model index.

In other words, in the limit of both infinite simulations
and infinite posterior draws per simulation SBC and binary prediction calibration for the model index have the same properties for
\(K = 2\). Nevertheless, the
performance of their practical implementations can differ when a finite
number of simulations is used.

When \(K \geq 2\), we can still run SBC for the model index and discover
many potential problems, although the equivalence between SBC and
prediction calibration will no longer hold. One relevant result from
\cite{modrak2025}~is that, in this case, changing the order of the submodels
can result in different \(\phi\) passing the SBC check. However, in
practice, examples when the result of an SBC check differs substantially after reordering a test quantity are hard to find.

The relationship between SBC and binary prediction calibration holds more generally --- the BMA supermodel is in the end just another Bayesian model, although one we typically cannot fit directly with MCMC. For any model $\pi$ over data space \(Y\) and parameter space
\(\Theta\) with \(f: \Theta \times Y \to \{0,1\}\) and for a candidate posterior \(\phi\)
we can define the random variable $D_{\phi,f}(y) = \mbox{Pr}_\phi(f(\theta, y) = 1 \mid y)$ and assuming the distribution of $D_{\phi,f}$ is not singular, $\phi$
passes continuous SBC w.r.t. \(f\) if and only if:
\begin{equation}
\forall p: \mbox{Pr}_\pi\left(f(\theta, y) = 1 \mid D_{\phi,f} = p\right) = p    
\end{equation}
 whenever the left-hand side is defined (i.e.~there is \(y\) in the
support of \(\pi_\text{marg}\) for which \(\phi\) predicts that the
expected value for \(f\) is \(p\)). See Theorem 1 in Appendix A for proof. 

Note that the $D_K$ variable defined in section~\ref{bayes-factors-and-binary-prediction-calibration}, is a special case of the $D_{\phi,f}$ variable defined here. Specifically, we have $D_k = D_{\phi_\text{BMA},g_{k}}$ where $g_{k}((i, \theta), y) = \mathbb{I}[i = k]$.

\subsection{Using posterior SBC with Bayes Factors}
To use posterior SBC with Bayes factors, 
we need to compute the implied
prior for the parameters of both models, but we also need the new prior
for the model index \(\bar\pi_\text{BMA}(i\mid y_1)\). Another important
factor in choosing \(y_1\) thus needs to be that it does not produce too
extreme implied prior probability for any of the models under
consideration.

Alternatively, we can tweak the original prior probability, to get
uniform distribution in the implied prior, i.e.~choose
\(\mbox{Pr}(\mathcal{M}_k)\) such that
\(\bar\pi_\text{BMA}(i\mid y_1) = \frac{1}{K}\). This is achieved by
setting
\begin{equation}
\mbox{Pr}(\mathcal{M}_k) \propto \prod_{i \neq k}\pi_\text{marg}^i(y_1).
\end{equation}
%

\subsection{Limitations of data-averaged posterior checking}

A fundamental limitation of data-averaged posterior checking is that it ignores the
mapping between data and posteriors --- there are many ways to change
predicted outcomes while keeping the average prediction identical. 
Assuming uniform prior over models, some
examples of specific model mixups that can never be detected by
data-averaged posterior checking include:

\begin{itemize}
\tightlist
\item
  \textbf{All models identical}, i.e. all fitted submodels are actually the same, even when data was simulated from substantially different submodels.
\item
  \textbf{Permuted model indices}, i.e., the fitted submodels are arbitrarily permuted. When
  \(K = 2\) this gives the inverse of the correct
  Bayes factor.
\item
  \textbf{Different predictor sets} in the context of regression models.
  Swapping any fitted submodel for one with different predictors than the one that was used to simulate the data will lead to identical data-averaged posterior of the model index as long as the implied prior distribution of
  the linear predictor is the same.
\end{itemize}

Note that all of those problems are not implausible to arise from a
simple programming error. 
The first two
problems will violate both prediction calibration and SBC (even without
extra test quantities) in all cases. The latter will almost certainly
fail those checks as well, although one may be able to construct some
contrived counterexamples. SBC with suitable test quantities will
discover all problems.

Additional problem is that a package may repeatedly assign very low
probability to the correct model and still pass data-averaged posterior checking by pushing all posterior model probabilities closer to extremes while maintaining the same average probability.

Moreover, for a binary
test quantity \(f : \Theta \times Y \to \{0, 1\}\) 
we have that whenever \(\phi\) passes SBC w.r.t. $f$ (which
is equivalent to having calibrated predictions for that quantity) it
will also have the correct data-averaged posterior w.r.t. \(f\).
See Theorem 3 in Appendix A. That is, in the limit of infinitely many simulations, data-averaged posterior checking provides
no benefit for Bayes factor validation over SBC or binary prediction calibration, although
it is still possible that its performance may be better with finitely many simulations.

\subsection{Implementing data-averaged posterior checking
}
As noted in Section~\ref{data-averaged-posterior}, the original implementation
of data-averaged posterior checking for Bayes factors \citep{schad_workflow_2023} 
suggests using one-sample Bayesian t-test on the distribution of posterior model probabilities.
This has both theoretical and practical limitations. 
On the theoretical side, we argue that control of false positives and power are primary concerns for a calibration test and those are naturally handled in a frequentist approach. Additionally,
there is no guarantee that the distribution of posterior model probabilities is close to normal --- it may be completely arbitrary.
So, strictly speaking, we require a test for
the mean of bounded variable with unknown distribution. Though our experiences is that in most
practical cases, the posterior model probabilities are well-behaved
enough for a classical one-sample t-test to work well.

In the rare cases, when the t-test approximation is problematic (e.g., when all observed
posterior probabilities are identical, or with a bimodal distribution
and low number of simulations), we would recommend using non-parametric
tests for the mean of bounded distribution of which we have found the test
of \citet{gaffke2005} to work best. The test is more thoroughly investigated by
\citet{learnedmiller2020cimean} ---
this test has been proven valid only in special circumstances, but no
counterexample is known, and it provides much larger power than the
alternatives --- this test is implemented in the \texttt{SBC} R package. 

A practical problem with the Bayesian t-test is that it requires us to set a free parameter (the prior standard deviation
of the alternative model). \citet{schad_workflow_2023} deal with this by reporting results for a range of prior choices which in our view makes the results harder to interpret, while providing no additional advantage.
We compare the empirical performance of all of the above mentioned tests in
Appendix C.

\subsection{Limitations of the Good check}

We immediately see that the
Good check shares some of the big limitations with data-averaged posterior checking
--- it also ignores the mapping between BFs and data and instead only
checks the overall distribution of BFs. The Good check conditions on
the true model, so there are two groups where averages could be explored, allowing for some
improvement over data-averaged posterior checking --- for example, the Good check can detect when we permute the two models (compute inverse of the Bayes factor). 
However, in the $k = 0$ version,
the Good check can never discover when the BF is computed in favor of a different model than assumed as long as it is distinct from $\mathcal{M}_1$.
In addition, the version
advocated by  \citet{Sekulovski2024goodcheck} involves substantial 
practical difficulties: we cannot make any assumptions about the distribution 
of the BFs (and their higher integer powers) and indeed those distributions can have a very large or even undefined variance, making
the sample mean a poor estimator of the expected value.
See Appendix D for examples of cases that pose such problems.

The absence of any bound on variance of the estimate explains the lack
of convergence for some cases and slow convergence for others 
(requiring tens of thousands of simulations) as
reported by \cite{Sekulovski2024goodcheck}.~It also prevents us from evaluating the
remaining uncertainty we have after running any finite number of simulations, as we cannot rule
out very rare, very large BF values. To be precise, since BFs have a lower bound of zero, we can in principle derive
useful lower confidence bounds for the mean (see e.g. \citealt{wang2003noparametriclower} and references therein), 
and thus seeing the sample mean $\hat{\mathbb{E}}(BF_{0,1} \mid  \mathcal{M}_1) \gg 1$ can be conclusive. 
On the other hand, no useful
upper confidence bound can exist without making further hard-to-verify assumptions and so observing $\hat{\mathbb{E}}(BF_{0,1} \mid  \mathcal{M}_1) \ll 1$ is by itself inconclusive. We note that
\citeauthor{Sekulovski2024goodcheck} do not make any attempt to formally control the error
rates of their procedure.

Interestingly, in both Examples 1 and 2 reported by \cite{Sekulovski2024goodcheck}, the simulation does not match the assumed model (some parameters are treated as constants instead of being drawn from the relevant prior) and thus the computed Bayes factors should be at least slightly incorrect. This mismatch is not picked up by the Good check in either case even though a large number of simulations is used. See Section~\ref{posterior-sbc-and-the-importance-of-using-correct-priors} for a similar example and how the mismatch is discovered with SBC.
Since the Good check as implemented  by \cite{Sekulovski2024goodcheck} fails to provide any means of error control, we will not evaluate this variant of the Good check further in this
paper.

\subsection{\texorpdfstring{Subsets of
\(Y\)}{Subsets of Y}}\label{subsets-of-y}

All of the checks we discuss in this paper also hold, if we focus only on a specific subset
\(\bar{Y} \subset Y\) of the data space:
SBC, and binary prediction calibration is directly
applicable to any subset of simulations that can be defined with just reference to the generated data (see Remark 4 in Appendix A; this is also the consequence of
Theorem 5 in \citealt{modrak2025}). This has two use cases: 1) we may want to
implicitly improve our prior by not fitting our models at all to
datasets that are unrealistic (i.e.,~doing rejection sampling from the
prior) and 2) we may want to investigate subsets of simulations if we
believe there are multiple different regimes that are worth
investigating (e.g.,~looking separately at fits that had problems
converging and those that converged). 
To keep the checks valid for Bayes factors,
the rejections of simulated datasets must be done for the whole BMA model, not when
simulating each model individually\footnote{The first author of this
  paper has spent an embarrassing amount of time tracking down the
  source of an apparent miscalibration in a model when the problem actually
  stemmed from rejecting datasets from each submodel separately. This
  illustrates the importance of trying to understand \emph{why} a model
  is miscalibrated before claiming there is a problem in computations.}.

Using data-averaged posterior checking for a subset of $Y$ is also possible but requires a little more care: we must account for the change in the implied prior
distribution. 
If we are unable to find an analytical form of
the prior conditioning on \(y \in \bar{Y}\) (or a sufficiently precise numerical value), we need to perform
data-averaged posterior checking with a two-sample test. For the model index,
we can use a paired two-sample test for equality of means. Similarly to the known prior case,
we cannot assume any specific distributional form, but in typical use cases a t-test provides 
a good enough approximation. The test of \citet{gaffke2005} can be applied as a paired test when t-test is unreliable.

\section{Evaluation methods}\label{evaluation-methods}

The fact that a check can recognize an incorrectly computed Bayes
factor/marginal likelihood
in the limit of
infinitely many simulations and infinitely precise computation does not
mean it can reliably recognize the problem with finite computational
resources. We run simulation studies to show the power of various
approaches to detect specific discrepancies.

We simulate \(i\) and then \(\theta_i\) and \(y\) from the corresponding
model. In all simulations we assume uniform prior over models. This is
without loss of generality as correct results depend on the correctness
of the Bayes factor computation, which in turn allows us to compute
results for any prior. 

To investigate the overall evolution of the various test statistics as the number of
simulations increase, we show full ``histories'' of the statistics. For short histories we compute the statistics after each simulation is added (step size of 1), for longer histories we increase the step size up to 200 simulations. 
To show how variable the evolution is, we always show 100 such histories. However, to keep computation costs manageable, we do not run 100 times more simulations, but instead sample simulations to be included in each history without replacement from a shared pool of simulations. The size of the pool is 2 -- 10 times the length of a single history.

For data-averaged posterior checking, we use the results from a standard
t-test (one-sample when the prior is known or paired two sample when prior is estimated from simulations). 
For binary prediction calibration, we use a bootstrap-based
miscalibration test based on score decomposition described in \citet{dimitraidis2021_reliability}. Specifically, miscalibration reported here is the
difference between the actual Brier score of the forecasts and Brier
score from a suitably recalibrated forecast which provides a meaningful
interpretation to its values. Following \citeauthor{dimitraidis2021_reliability}~we calculate
bootstrap-based null distribution for miscalibration by repeatedly
sampling new observed binary outcomes as if the computed probabilities
were correct. 

For SBC, we use the $\gamma$ statistic of \citet{sailynoja_graphical_2021}, which is also used in \citet{modrak2025}. We have $\gamma = 2 \min_{i\in \{1, \dots, M+1\}}\left(\min\{\text{Bin}(R_i | S, z_i), 1 - \text{Bin}(R_i - 1 | S, z_i)\}\right)$ where $M$ is the number of draws in the sample obtained from the posterior, $S$ is the number of simulations, $z_i = \frac{i}{M + 1}$ is the expected proportion of observed ranks smaller than $i$, $R_i$ is the observed count of ranks smaller than $i$, and $\text{Bin}$ is the CDF of the binomial distribution.
\citeauthor{sailynoja_graphical_2021} report a method to evaluate quantiles of the
distribution under the null, but no efficient method to evaluate its
CDF is available and so we cannot directly compute a $p$-value. Instead we
report the log of the ratio of the statistic, to its 5th percentile
under the null (i.e.~log ratio \(< 0 \implies p < 0.05\)). To help
interpretation when SBC checks pass, we also report the maximal absolute
difference between uniform CDF and empirical CDF the tests would detect
(which is the difference at the middle of the rank distribution).

\section{Toy examples}\label{toy-examples}
We first evaluate the proposed methods on two toy examples.

\subsection{Single binary
observation}\label{single-binary-observation}

Let us start with a comparison of two extremely simple models
with no parameters (\(\Theta_0 = \Theta_1 = \emptyset\)):
\begin{gather}
Y = \{0,1\}, \ \ \mbox{Pr}(y = 1 \mid  \mathcal{M}_0) = \frac{1}{5}, \ \  \mbox{Pr}(y = 1 \mid  \mathcal{M}_1) = \frac{4}{5} \notag \\
\mbox{Pr}(\mathcal{M}_0) = \mbox{Pr}(\mathcal{M}_1) = \frac{1}{2}
\end{gather}
Any candidate posterior can be described just by two numbers
\(b_0 = \phi_\text{BMA}(\mathcal{M}_1 \mid  Y = 0)\) and
\(b_1 = \phi_\text{BMA}(\mathcal{M}_1 \mid  Y = 1)\). Analytically, we
can show that \(\phi_\text{BMA}\) will satisfy data-averaged
posterior checking if and only if \(b_0 + b_1 = 1\), i.e.~allowing
infinitely many wrong Bayes factors, including assigning 100\%
probability to the wrong model or having
\(\phi_\text{BMA}(\mathcal{M}_1 \mid  Y = 0) = 1 - \pi_\text{BMA}( \mathcal{M}_1 \mid  Y = 0)\).
To be specific data-averaged posterior checking for this model implies:
\begin{equation}
\forall i \in \{0,1\}: \mbox{Pr}(\mathcal{M}_i) = \sum_{y=0}^1  \phi_\text{BMA}(i \mid  Y = y) \sum_{k=0}^1 \mbox{Pr}(y \mid  \mathcal{M}_k) \mbox{Pr}(\mathcal{M}_k)
\end{equation}
Averaged over $i$, both values of \(y\) are equally
likely,
\(\sum_{k=0}^1 \mbox{Pr}(y \mid  \mathcal{M}_k) \mbox{Pr}(\mathcal{M}_k) = \frac{1}{2}\),
\begin{equation}    
\frac{1}{2} = \frac{1}{2} \sum_{y=0}^1  b_y \wedge
\frac{1}{2} = \frac{1}{2}\sum_{y=0}^1  (1 - b_y),
\end{equation}
where both equations reduce to \(b_0 + b_1 = 1\).

\begin{figure}[b!]

\centering{

\includegraphics[width=\figurescale\linewidth]{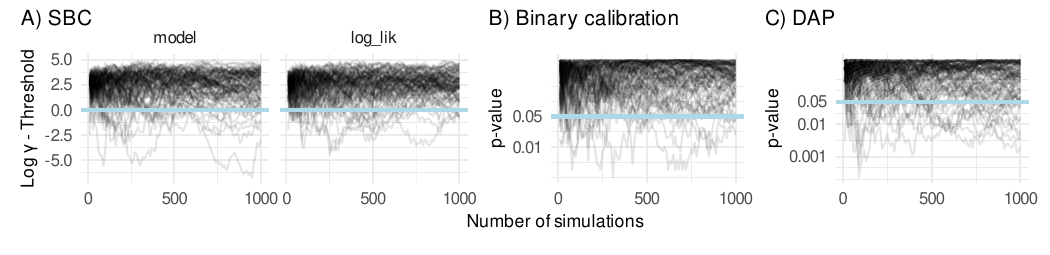}

}
\caption{\label{fig-binary-correct}\small Histories of check statistics for the binary model when the result is correct. \historyinfoblue\ We see that in
all cases, violations tend to be non-severe and short-lived. A) shows the
log-gamma statistic of SBC for the model index and log likelihood, B) is the p-value of the
bootstrapped miscalibration test of \citet{dimitraidis2021_reliability}~and C)  is the
p-value from t-test for data-averaged posterior (DAP). 
Note that the p-values from the miscalibration
test are capped at \(\frac{1}{2000}\) due to the number of bootstrap
samples used.}

\end{figure}

In contrast, \(\phi_\text{BMA}\) will pass the prediction calibration
check (and hence SBC for \(f(i, y) = i\)) only for the correct posterior
and for the prior (\(b_i = \mbox{Pr}(\mathcal{M}_i) = \frac{1}{2}\)).
The proof is that either \(b_0 = b_1\) and then the model is calibrated
only if \(b_0 = b_1 = \frac{1}{2}\) or that \(b_0 \neq b_1\) and then
the model is calibrated only when
\(b_0 = \frac{1}{5}, b_1 = \frac{4}{5}\).
Additionally, checking SBC w.r.t. the likelihood
\(g(i, y) = a_i^y (1 - a_i)^{1-y}\) also rules out the
\(b_i = \frac{1}{2}\) option.
How quickly do we find problems empirically? When the
model is correct we find no big problems --- the various test statistics
have the expected random behavior and only rarely drop below the
relevant thresholds (\refextrafigure{fig-binary-correct}).

\begin{figure}[t]

\centering{

\includegraphics[width=\figurescale\linewidth]{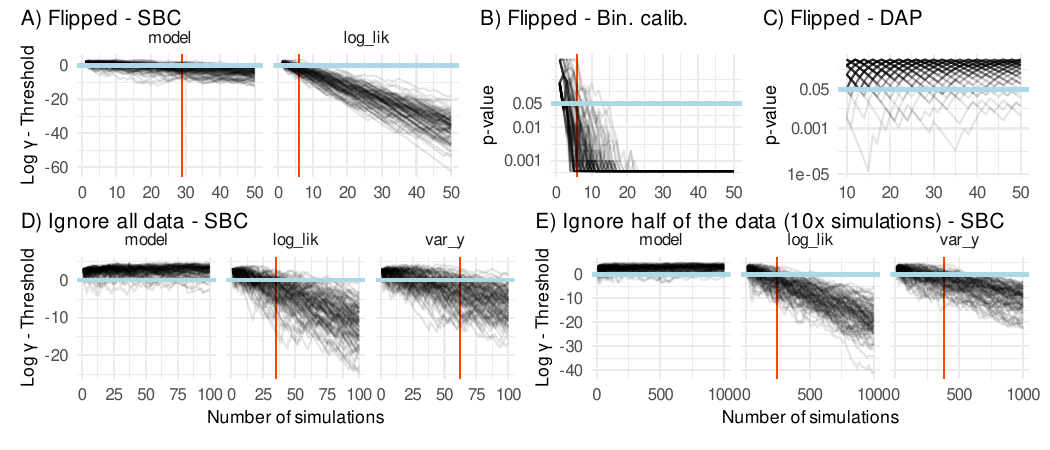}

}
\caption{\label{fig-flipped-pnb-ignore}\small Histories of check statistics for the flipped binary model (A - C), Poisson-NB model ignoring all data in BF computation (D) and ignoring half of the data (E). For the Poisson-NB model, SBC statistics are reported for the model index
and two data-dependent derived quantities: the log likelihood (log\_lik)
and variance estimate (var\_y).  \historyinfoblue\ A), D) and E) show the
log-gamma statistic of SBC for the model index and log likelihood, B) is the p-value of the
bootstrapped miscalibration test of \citet{dimitraidis2021_reliability}~and C) is the
p-value from t-test for data-averaged posterior (DAP). 
Note that the p-values from the miscalibration
test are capped at \(\frac{1}{2000}\) due to the number of bootstrap
samples used.}

\end{figure}%

However, if we compute the Bayes factors as if the probabilities were
flipped (i.e.,~computing the inverse of the true Bayes factor), the
problem is very quickly flagged by all tests except for data-averaged
posterior checking (Figure~\ref{fig-flipped-pnb-ignore} A - C). This model is however
somewhat artificial as it has only two possible Bayes factor values. Let
us move to a slightly more interesting example.

\subsection{Poisson vs.~negative
binomial}\label{poisson-vs.-negative-binomial}

To provide a slightly richer example, we used the following model:
\begin{equation}
\begin{split}
\begin{aligned}
\mathcal{M}_0: y_1,\dots,y_N &\sim \text{Poisson}(3)\\
\mathcal{M}_1: y_1,\dots,y_N &\sim \text{NB}_2(3, 5)
\end{aligned} 
& \hspace{4em}
\mbox{Pr}(\mathcal{M}_0) = \mbox{Pr}(\mathcal{M}_1) = \frac{1}{2}
\end{split}
\end{equation}
In this setting, the correct BF is just the likelihood ratio of the two
models. In our experiments we used \(N = 25\) as this makes the observed
posterior probabilities of models values cover the full \([0,1]\)
interval.
%
%
%
Beyond the model index, we use two derived quantities in SBC, one is the
log-likelihood, second is an estimate of variance, specifically
\(f(0, y) = \text{Mean}(y)\), $f(1, y) =  \text{Var}(y)$.
This is an example of a simple data-dependent test quantity designed
with the specific models in mind: under $\mathcal{M}_0$ (Poisson), the mean should be equal to
the variance. This variance-based quantity illustrates that one does
not need to bring in the full model likelihood (which might be laborious
to re-implement) to gain benefits of data-dependent quantities.

Let us start with the case that the posterior probabilities are all
\(\frac{1}{2}\) --- we know from theory that neither data-averaged posterior checking, nor
binary prediction calibration can detect this regardless of the number of simulations 
(and that indeed holds empirically in our experiments, results not shown here).
In contrast, SBC with data-dependent quantities
typically discovers the problem reasonably quickly (see
Figure~\ref{fig-flipped-pnb-ignore} D).

One may argue that completely ignoring data in the computation is easy
to discover, simply check for constant BFs. So the next step is a
situation where just half of the datapoints are ignored --- this still
cannot be detected by binary prediction calibration or data-averaged
posterior checking, but SBC will detect it with suitable derived
quantities. The posterior is now closer to correct, so we require more
simulations to discover the problem, but both data-dependent quantities
uncover the issue (see Figure~\ref{fig-flipped-pnb-ignore} E).

\begin{figure}[t!]
\centering{

\includegraphics[width=\figurescale\linewidth]{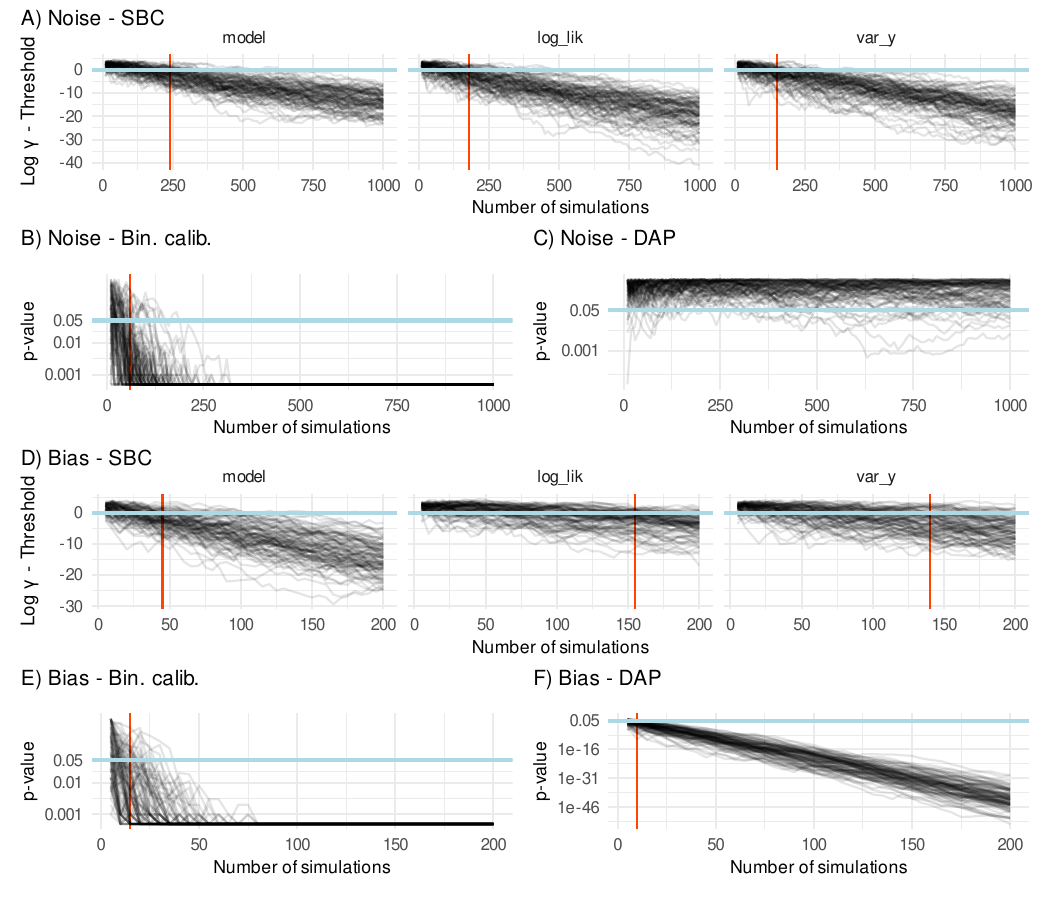}

}

\caption{\label{fig-pnb-noise-bias}\small Histories of check statistics for
oisson-NB model with noise in BF (A - C) and with bias in BF (D - F).  A) and D) show the log-gamma statistic of
default SBC, B) and E) is the p-value of the bootstrapped miscalibration test
of \citet{dimitraidis2021_reliability}~and C) and F) is the p-value from t-test for
data-averaged posterior checking (DAP).  \historyinfo  
We see both problems quickly detected by miscalibration test and SBC, while DAP does not detect noise but is the fastest to detect bias.
}

\end{figure}

Now let us try to introduce noise into the BF, specifically we will add
normal noise with standard deviation 2 to the logarithm of the BF. This is relatively
quickly discovered by binary calibration and SBC, but generally missed
by data-averaged posterior checking (see Figure~\ref{fig-pnb-noise-bias} A-C).

In our final toy scenario, we add 2 to the logarithm of the Bayes factor for all
simulations, mimicking the use of a wrong normalization constant in one of the models. This is discovered quickly by all of the
options, but data-averaged posterior checking requires the fewest simulations (see Figure~\ref{fig-pnb-noise-bias} D-F).

\subsection{Summary of toy examples}\label{summary-of-toy-examples}

In more general models, SBC with any realistic number and choice of test
quantities may permit many wrong posteriors. However, when using
suitable data-dependent test quantities, those wrong posteriors are hard
to construct and are unlikely to be the result of a bug in the
computation. On the other hand, wrong posteriors satisfying
data-averaged posterior checking are easy to come by and can plausibly
arise due to a programming error (e.g.,~flipping the model indices, not
including a part of the data) or bad numerical approximations (increased
variance, but no bias).

\section{Realistic examples}\label{realistic-examples}


Now we focus attention on more realistic datasets. First is the \textit{turtles} example from the
documentation of the \texttt{bridgesampling} package \citep{bridgesampling_package}.
Model  $\mathcal{M}_1$ is a probit regression model with a single varying
intercept:
\begin{equation}
\begin{aligned}
y_i \mid \alpha, \beta, \mathbb{\gamma} &\sim  \text{Bernoulli}(\Phi(\alpha + \beta x_{i} + \gamma_{g_i})),  &i = 1,2,\ldots,N\\
\gamma_j \mid \tau &\sim N(0, \tau),  &j = 1,2,\ldots, G \\
\alpha &\sim N(0, \sqrt{10}), \:
\beta \sim N(0, \sqrt{10}),  \: &\pi_\text{prior}(\tau^2) = \left(1 + \tau^2\right)^{-2},
\end{aligned}    
\end{equation}
where we have $N$ observations split into $G$ groups ($g_i$ is the group for the $i$-th observation) and \(\Phi\) is the normal cumulative distribution function. 
Model $\mathcal{M}_0$ is then the same but with the varying intercept removed,
i.e.,~we set \(\tau = 0\). Here, and throughout this paper we parametrize the normal distribution with mean and sd.

We perform two checks --- for one we use standard SBC with
\(N = 70, G = 10\) (this keeps the distribution of observed BFs
reasonably wide) 
and simulate \(x_i \sim N(0, \frac{1}{3})\). 
This tends to produce a
substantial number of datasets where \(y_i\) has zero or very little
variance. This can be overcome by rejection sampling (see section~\ref{subsets-of-y}) --- 
we remove all datasets (and the associated prior
draws) where \(\text{Avg}(y) < \frac{1}{10}\) or
\(\text{Avg}(y) > \frac{9}{10}\). In this example, rejections change the
implied prior to \(P(\mathcal{M}_1) \simeq 0.6\).  After running 2 000 simulations
we find no problems (see \refextratable{tab-successful-checks}).

For the second check, we focus the simulations to the neighborhood of
the actual data by means of posterior SBC. 
To this end, we pick data for half of the groups as
\(y_1\), fit both models to this smaller dataset, and calculate the
posterior model probabilities. We then use the fitted parameters to
simulate data for the remaining groups.

We could use the implied model probability as prior for the model index,
but to show an alternative, we can also choose \(P(M_1)\) such that 
\(P(M_1\mid y_1) = \frac{1}{2}\) and so our simulations
will be more balanced. In this example, this implies \(P(M_1) \simeq 0.77\). 
This new \(P(M_1)\) is then used when
converting the computed Bayes factor into posterior model probability. After running 2 000 simulations we see no
issues (see \refextratable{tab-successful-checks}).

\subsection{Discovering bad normalization
constants}\label{discovering-bad-normalization-constants}

Recently \citet{Tsukamura2024_normalization}
have examined the prevalence of missing or incorrect normalizing
constants for posterior densities. The authors note that 9 out of 21
investigated Stan models using a truncated distribution neglected the
normalizing constant. If those models were used with
\texttt{bridgesampling}, the resulting marginal likelihoods (and hence
Bayes factors) would be biased. Here we test how quickly our checks
discover this problem. We use the same model as in the previous section,
except for a more standard prior on random effect standard deviation
\(\tau \sim \text{HalfN}(0, 1)\). We also neglect the normalization
constant and so the density and marginal likelihood is wrong by a factor
of 2.
This is quickly picked up by all checks, with binary prediction
calibration and data-averaged posterior checking being the fastest
(Figure~\ref{fig-turtles-bad-norm-ranef} A-C).

\subsection{Posterior SBC and the importance of using correct
priors}\label{posterior-sbc-and-the-importance-of-using-correct-priors}

When our models use improper priors for some parameters, it may be
tempting to use a fixed value or draw the affected parameters from some
surrogate prior distribution. In this case however, the assumptions of
all of the consistency criteria do not hold. And indeed, we see
violations of all checks.

Problems are largest when a parameter with improper prior interacts with
some parameter of actual interest. A practically highly relevant example
is the interaction of the prior on the intercept (which is often taken
as improper) with the prior on random effects, see \citet{ogle2020_identifiability_mixed} for more
discussion of this phenomenon.

We test a linear regression model with a single binary predictor and a
single random effect. We will be using the \texttt{BayesFactor} package \citep{bayesfactor_package},
which parametrizes the model as:
\begin{equation}
\begin{aligned}
y_i \mid \alpha, \beta, \mathbb{\gamma}, \sigma &\sim  N(\alpha + \beta x_{i} + \gamma_{g_i}, \sigma),  &i = 1,2,\ldots,N\\
\gamma_j \mid \sigma,\tau &\sim N\left(0, \frac{\sqrt{2}}{4}\sigma \sqrt{\tau} \right), &j = 1,2,\ldots, G \\
\end{aligned}
\end{equation}
and assumes the following prior:
\begin{equation}
\begin{aligned}
\tau &\sim \text{InvGamma}\left(\frac{1}{2},\frac{1}{2} \right), &
\beta \mid \sigma &\sim N\left(0, \frac{\sqrt{2}}{4} \sigma \right), & \pi_\text{prior}(\alpha, \sigma^2) &= \frac{1}{\sigma^2}.
\end{aligned}
\end{equation}
Where the binary predictor $x_i$ is coded as \(x_i = \pm \frac{\sqrt{2}}{2}\). We compared this
to a model without random effect, i.e.~\(\tau = 0\). The
implied prior on \(\gamma_j\) is multivariate Cauchy and the prior on
\(\alpha, \sigma^2\) is improper and cannot be sampled.

\begin{figure}[t!]

\centering{

\includegraphics[width=\figurescale\linewidth]{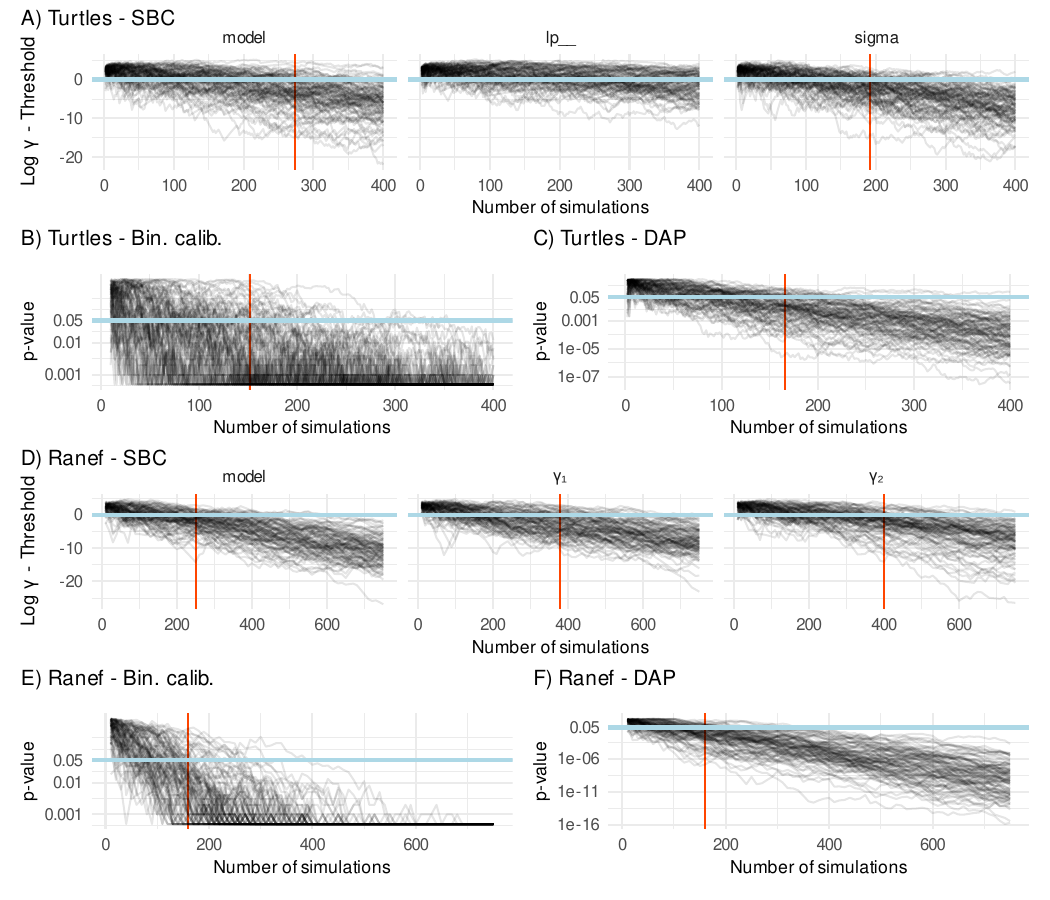}

}

\caption{\label{fig-turtles-bad-norm-ranef}\small Histories of check statistics for
bad normalization constant in the turtles model (A - C) and random effect model using constant intercept and standard deviation (D - F). A) and D) show the log-gamma
statistic of SBC, (lp\_ = unnormalized log-posterior density), B) and E) is the p-value of the bootstrapped
miscalibration test~and C) and F) is the p-value from
t-test for data-averaged posterior checking.  \historyinfo}
\end{figure}

\begin{figure}

\centering{

\includegraphics[width=\figurescale\linewidth]{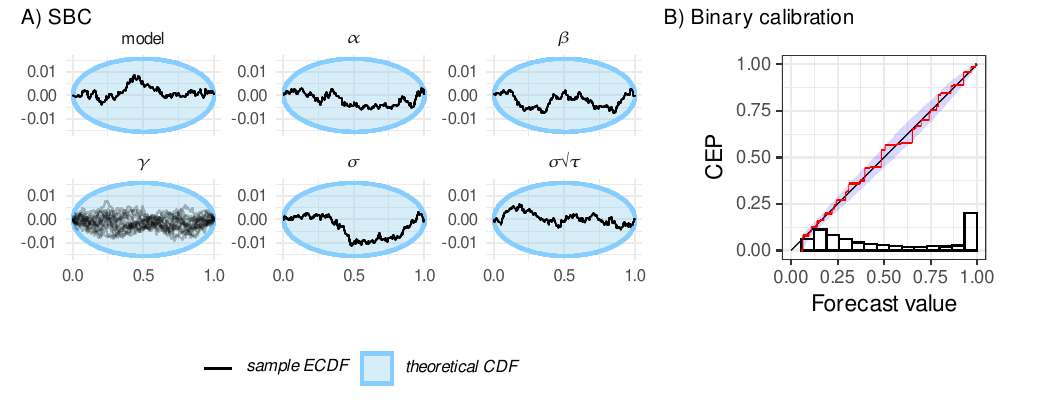}

}
\caption{\label{fig-ranef-post}\small Result of calibration checks for random
effect model via the \texttt{BayesFactor} package, using posterior SBC. A) empirical CDF difference plots, 
B) binary calibration plot, \calibrationplotinfo }

\end{figure}

If we choose the overall intercept and residual standard deviation as
fixed constants (here \(\alpha = 0, \sigma = 1\)) we obtain an apparent
miscalibration for both the model index and all the individual random
effect parameters (see Figure~\ref{fig-turtles-bad-norm-ranef} D-F). In this case
the problem is the most quickly picked up by the binary prediction
calibration test. Data-averaged posterior checking and SBC for the model
index follow quickly. However this is a false positive due to
the mismatch between the simulator (assuming constant
\(\alpha, \sigma\)) and the model. Employing posterior SBC using a
dataset with 4 groups and 3 observations each as \(y_1\) (to make all
parameters well identified for sampling) produces a proper prior for the
intercept and standard deviations and results in perfect calibration in
all methods even after running 10 000 simulations
(\refextrafigure{fig-ranef-post} and Table~\extraref{tab-successful-checks}). 

The previous example is the simplest setting where we
obtained apparent strong miscalibration due to improper priors for shared
parameters, however, the problem subtly affects even much simpler
models -- in Appendix F we show a similar problem with Bayesian t-test as implemented in the
\texttt{ttestBF} function in the \texttt{BayesFactor} package.

\subsection{Multiple models}\label{multiple-models}

Finally, we test a comparison between multiple nested linear mixed models ($K = 4$). The most complex model  \(\mathcal{M}_3\) is

\begin{equation}
\begin{aligned}
y_i \mid \alpha, \beta, \mathbb{\gamma}, \sigma &\sim  N(\alpha + \beta_1 x_{1,i} + \beta_2 x_{1,i} + \gamma_{g_i}, \sigma),  &i = 1,2,\ldots,N\\
\gamma_j \mid \tau &\sim N\left(0, \tau \right), &j = 1,2,\ldots, G \\
\end{aligned}
\end{equation}
and assumes the following prior:
\begin{equation}
\begin{aligned}
\alpha &\sim N\left(2, 2 \right), & \beta &\sim N\left(0, \frac{1}{2} \right), & \tau &\sim \text{Gamma}\left(8,4 \right), & \sigma \sim \text{HalfN}(0, 3)
.
\end{aligned}
\end{equation}
We obtain  \(\mathcal{M}_2\) by assuming $\beta_1 = 0$, \(\mathcal{M}_1\) by assuming $\beta_2 = 0$ and \(\mathcal{M}_0\) by assuming $\beta_1 = \beta_2 = 0$. In our simulations we have $N = 60$, $G = 15$ and we fit the model with the \texttt{brms} package \citep{brms}. For Bayes factor computation, we consider bridge sampling and the Savage-Dickey density ratio as implemented in the \texttt{hypothesis} function in \texttt{brms}. This implementation does not support conjunctions of hypotheses on multiple parameters, so we directly obtain $BF_{13}, BF_{23}$ from the $\mathcal{M}_3$ fit and then obtain $BF_{01}$ from $\mathcal{M}_1$ fit.  

We ran 40 000 simulations, though we omitted 835 from final results due to errors in bridge sampling or other problems in computation. Bridge sampling shows no problems and the Savage-Dickey estimates are also quite well behaved and differ noticeably from bridge sampling results only for Bayes factors below $0.01$ where the Savage-Dickey estimates are lower (see Appendix G). We see only a mild and inconclusive violation of the DAP criteria for the indicator variables for $\mathcal{M}_3$(95\% CI for difference: [-0.0031, -0.0002], p = 0.027) and no problem for SBC or binary calibration. 

We then introduce two types of stronger problems --- one local and one global. For the former, we emulate a creative but lazy user who wants to compute all of the Bayes factors just from the $\mathcal{M}_3$ fit. Specifically, we compute $BF_{03}$ by evaluating the hypothesis $\beta^2_1 + \beta_2^2 = 0$ which is  equivalent to $\mathcal{M}_0$, but the \texttt{brms} implementation assumes that the tested point hypothesis does not lie on the boundary of its support and this introduces a positive bias into $BF_{03}$. We call this scenario ``trick''. For the latter, we reduce the number of draws for each fit to 30 and call this scenario ``few draws'', increasing noise in all of the BF estimates (see Appendix G for direct comparison of the results).

\begin{figure}

\centering{

\includegraphics[width=\figurescale\linewidth]{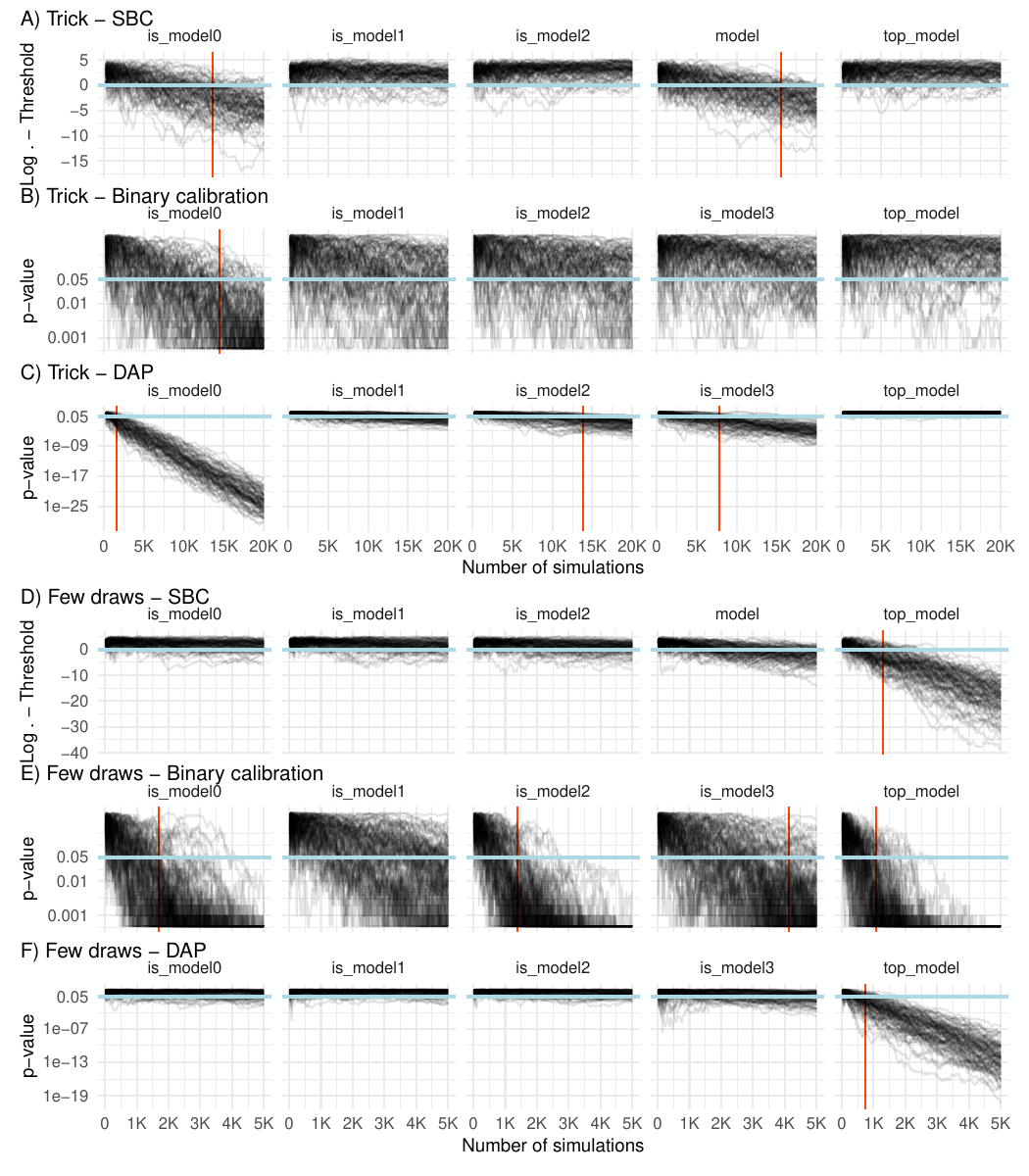}

}

\caption{\label{fig-multiple-models} \small Histories of check statistics for
the multiple models scenario.
A) and D) show the log-gamma
statistic of default SBC, B) and E) is the p-value of the bootstrapped
miscalibration test of \citet{dimitraidis2021_reliability}~and C) and F) is the p-value from
t-test for data-averaged posterior.  The results are shown for binary indicators of the individual models and the top model. For SBC we do not show the indicator for $\mathcal{M}_3$ (which looks similar to the $\mathcal{M}_2$ indicator in both cases) and show the four-valued model index instead.}

\end{figure}

In the trick scenario, DAP for the $\mathcal{M}_0$ indicator discovers the problem fastest, while DAP for other quantities, SBC and binary calibration require substantially more simulations. The top model indicator shows no problems at all (see \refextrafigure{fig-multiple-models} A--C). In the few draws scenario the top model indicator is the most sensitive, especially for DAP where it discovers the problem first, followed by SBC and binary calibration for top model and then binary calibration for the other model indices (see \refextrafigure{fig-multiple-models} D--F). This highlights that both individual model indicators as well as the top model indicator have their strengths -- the former are more sensitive to local problems and the latter to global problems. For SBC, the four-valued model index was in all cases at least as sensitive as the best of the binary model indicators.

To demonstrate scalability, Appendix H presents a use case with $2^{100}$ competing models where the choice of model is explicitly encoded with 100 discrete variables in a JAGS \citep{plummer2003jags} linear regression model. The main result is that we validate that JAGS is able to sample the model reliably and that when we introduce errors by shuffling some of the indicator variables, the problem is detected in 2 to 3 simulations by SBC for the log likelihood of the model. SBC for a suitable ordering over the models or taking the minimal p-value (after multiplicity adjustment) across tests for individual indicator can also perform reasonably well. The use case also shows that our methods apply even when marginal likelihoods/Bayes factors are not explicitly computed during model selection.

To summarize, when the number of models is relatively small, we recommend to check all of the individual model indicators, the top model indicator and the integer-valued model index. When the number of models makes this unfeasible, we need to aggregate the information. A good option is using the log likelihood as a test quantity.

\section{Discussion}\label{discussion}

We have provided new theoretical insights and empirical evaluation of multiple methods 
for checking the correctness of Bayes factor computation. All of the methods we discuss in this paper are available for easy use
with the \texttt{SBC} R package \citep{SBC_package}.

\subsection{Validating Bayes factor computation}
We have shown that SBC and binary prediction calibration are viable
methods to assess computation of Bayes factors. Beyond detecting failures
in computation with controlled error rates, 
both methods also report remaining uncertainty
we have about possible miscalibration that would not be detected with a given 
simulation budget. The Good check as implemented by \cite{Sekulovski2024goodcheck} provides no error control and is in our view
rarely useful, though other iterations of the idea could still prove their merit.

We have shown that whenever data-averaged posterior checking signals a problem in Bayes factor computation,
the problem will also be found with SBC or binary prediction calibration. 
At the same time, data-averaged posterior checking fails catastrophically
in many examples where SBC and binary prediction calibration work.
On the other hand, there are scenarios where data-averaged posterior checking  
detects problems with fewer simulations than SBC/binary prediction calibration. 
When the BFs are just very mildly biased (as when using Savage-Dickey density ratio in Section~\ref{multiple-models}),
the difference can be substantial. Using
data-averaged posterior checking for validation of Bayes factor computation
should thus be considered alongside SBC or binary
prediction calibration. We recommend the standard t-test for testing data-averaged posterior in typical cases. 
If there are very few simulations and the distribution of posterior model probabilities is far from 
approximately normal, we recommend the test of \citet{gaffke2005}.

We have shown that in theory, binary prediction calibration is a special case
of SBC. However, in the examples we ran, binary prediction calibration checks 
via the miscalibration test of \cite{dimitraidis2021_reliability} tended to have
slightly higher power to detect discrepancies than SBC for the model
index and sometimes even than SBC for the log likelihood. This is most
likely due to SBC throwing away information via random tiebreaking for
discrete variables (see Algorithm~\ref{sbcalg})--- increasing power of SBC in those settings is an
ongoing research project of ours. We consider it out of scope of this
paper and leave it as future work. For checking the calibration of binary variables, 
we would currently recommend binary prediction calibration over SBC.

However, unlike binary prediction calibration, SBC can be run for any test
quantity. We have observed several issues that binary prediction calibration missed
that were visible when using SBC with data-dependent test
quantities: (partially) ignoring data (Section~\ref{poisson-vs.-negative-binomial}) a
nd using incorrect prior for the variance in Bayesian t-test (Appendix F). 
So we argue that binary prediction calibration for the model
index should be accompanied by SBC for other model parameters and
suitable data-dependent test quantities.
Another advantage of running SBC for the full BMA model is that we can
use the same simulations to check that the individual models are well
implemented/calibrated --- if the individual models are wrong, we cannot
make any strong inferences about the method for computing Bayes factors.

Our results also show that to rule out problems in Bayes factor
computation, we need to use a lot of simulations ---
even quite strong miscalibration can require several hundred simulations
to be reliably uncovered (e.g., Figure~\ref{fig-flipped-pnb-ignore}E).

\subsection{Potential issues and their remedies}
Using multiple test quantities or even multiple methods for checking the
calibration of the model leads to multiple testing issues. However,
here Type II error (false negatives) is the biggest problem and the results for 
different methods/quantities are positively correlated, reducing the usefulness 
of a formal multiplicity correction. 
If a problem with the computation is suspected, we can
usually run more simulations to discern flukes from real problems with
reasonable expense. Additionally, in a perfect world, a failed
calibration check is followed by investigation to get understanding why
the computation failed, reducing false positives.

Seeing some of the proposed tests fail should not immediately lead to
claims of computational problems. 
By far the most common source of apparent miscalibration we have seen
while developing this paper stemmed from mismatches between simulation
code and the models to be compared, not from failure of computation. It follows that
packages that allow us to use the same model formulation for simulation,
fitting and (for posterior SBC) prediction (e.g., \texttt{brms}; \cite{brms}) make
checking Bayes factor computations much easier as this source of
problems can be ruled out.

A practical problem we encountered with checking the methods in
\texttt{BayesFactor} package was that the priors in use are described only in
research papers and require non-trivial effort to fully reconstruct. In
contrast, when \texttt{brms} \citep{brms} or a Stan program \citep{stan} was used to compute the
models the priors are transparently accessible.

Faced with miscalibrated Bayes factors, it may be tempting to use the
simulation results to recalibrate the method by transforming the posterior model 
probability with the inverse of the observed calibration curve. This would make the 
probabilities exactly calibrated across the simulations we ran, however, recalibration
with SBC does not work in general \citep{gelman2023recalibration}
so we caution against this approach.

\subsection{Practical recommendations}
Our recommendation for researchers who wish to implement and/or evaluate
tools that compute the Bayes factor is to use SBC for the full BMA model
accompanied by a binary calibration check for the model index and
data-averaged posterior checking. At least several hundred simulations per
validation scenario should be run to have reasonable power. If high precision is required, thousands of simulations should
be run. The required number of simulations to detect various discrepancies we observed in this work is summarized in \refextratable{tab-power}.

In all tests we ran, as well as those tested by
\cite{schad_workflow_2023, schad_aggregation_2024, schad2025bf},
both \texttt{bridgesampling} \citep{bridgesampling_package} and the \texttt{BayesFactor} \citep{bayesfactor_package}
packages provided reliable Bayes factor estimates or a warning on inaccuracy.
We therefore currently believe that for low-stakes statistical workflow
involving Bayes factors and standard models, it is safe to rely on those
packages without further testing. However, when using a less standard
model or when the costs of being wrong are non-trivial, we encourage
running at least several hundred simulations of SBC, preferably tailored
to the dataset in question with posterior SBC.

Our final recommendation concerns prior choice. 
While it is well known that Bayes factors are very sensitive to priors, prior
choice tends to focus on the parameters corresponding to the effect of treatment or other
quantity of interest. Priors for parameters that are shared across all models and not
of interest themselves (e.g., intercepts, standard deviations) are given less thought or
even taken as improper.
The fact that simulating from a mismatched prior for those shared parameters can lead to noticeably miscalibrated Bayes factors (Section~\ref{posterior-sbc-and-the-importance-of-using-correct-priors})
 implies that those priors may influence the resulting BFs considerably. 
This can be interpreted as either a call for greater care when setting priors or
as another evidence of the brittleness of BFs.

\begin{acks}[Acknowledgments]
We thank Eric-Jan Wagenmakers and Nikola Sekulovski for helpful comments and suggestions on an earlier version of this manuscript.
\end{acks}

\bibliographystyle{ba}
\bibliography{main}

@article{modrak2025,
author = {Martin Modr{\'a}k and Angie H. Moon and Shinyoung Kim and Paul B{\"u}rkner and Niko Huurre and Kateřina Faltejskov{\'a} and Andrew Gelman and Aki Vehtari},
title = {{Simulation-Based Calibration Checking for Bayesian Computation: The Choice of Test Quantities Shapes Sensitivity}},
volume = {20},
journal = {Bayesian Analysis},
number = {2},
publisher = {International Society for Bayesian Analysis},
pages = {461 -- 488},
year = {2025},
doi = {10.1214/23-BA1404},
URL = {https://doi.org/10.1214/23-BA1404}
}

@inproceedings{radev2023bfsbi,
  title = {Jana: {{Jointly}} Amortized Neural Approximation of Complex {{Bayesian}} Models},
  shorttitle = {Jana},
  booktitle = {{{Uncertainty}} in {{Artificial Intelligence}}},
  author = {Radev, Stefan T. and Schmitt, Marvin and Pratz, Valentin and Picchini, Umberto and K{\"o}the, Ullrich and B{\"u}rkner, Paul-Christian},
  year = 2023,
  pages = {1695--1706},
  publisher = {PMLR},
}

@article{aguilar2025generalized,
  title={Generalized Decomposition Priors on R2},
  author={Aguilar, Javier Enrique and B{\"u}rkner, Paul-Christian},
  journal={Bayesian Analysis},
  volume={1},
  number={1},
  pages={1--34},
  year={2025},
  publisher={International Society for Bayesian Analysis},
  doi = {10.1214/25-BA1524}
}

@article{meng_warp_2002,
	title = {Warp bridge sampling},
	volume = {11},
	doi = {10.1198/106186002457},
	number = {3},
	journal = {Journal of Computational and Graphical Statistics},
	author = {Meng, Xiao-Li and Schilling, Stephen},
	year = {2002},
	pages = {552--586},
}

@misc{schad2025bf,
      title={How accurate are Bayes factor-based null hypothesis tests? A simulation study}, 
      author={Daniel J. Schad and Martin Modrák and Shravan Vasishth},
      year={2025},
      eprint={2406.08022},
      archivePrefix={arXiv},
      primaryClass={stat.ME},
      url={https://arxiv.org/abs/2406.08022}, 
}

@Article{Sekulovski2024goodcheck,
author="Sekulovski, Nikola
and Marsman, Maarten
and Wagenmakers, Eric-Jan",
title="A Good check on the Bayes factor",
journal="Behavior Research Methods",
year="2024",
month="Dec",
day="01",
volume="56",
number="8",
pages="8552--8566",
issn="1554-3528",
doi="10.3758/s13428-024-02491-4",
url="https://doi.org/10.3758/s13428-024-02491-4"
}

@article{schad_aggregation_2024,
	title = {Data aggregation can lead to biased inferences in {Bayesian} linear mixed models and {Bayesian} analysis of variance.},
	issn = {1939-1463, 1082-989X},
	url = {https://doi.apa.org/doi/10.1037/met0000621},
	doi = {10.1037/met0000621},
	language = {en},
	urldate = {2025-06-02},
	journal = {Psychological Methods},
	author = {Schad, Daniel J. and Nicenboim, Bruno and Vasishth, Shravan},
	month = jan,
	year = {2024},
}

@article{schad_workflow_2023,
	title = {Workflow techniques for the robust use of bayes factors.},
	volume = {28},
	issn = {1939-1463, 1082-989X},
	url = {https://doi.apa.org/doi/10.1037/met0000472},
	doi = {10.1037/met0000472},
	language = {en},
	number = {6},
	urldate = {2025-06-02},
	journal = {Psychological Methods},
	author = {Schad, Daniel J. and Nicenboim, Bruno and Bürkner, Paul-Christian and Betancourt, Michael and Vasishth, Shravan},
	month = dec,
	year = {2023},
	pages = {1404--1426},
}

@misc{fazio2025primedpriorssimulationbasedvalidation,
      title={Primed Priors for Simulation-Based Validation of Bayesian Models}, 
      author={Luna Fazio and Maximilian Scholz and Javier Enrique Aguilar and Paul-Christian Bürkner},
      year={2025},
      eprint={2408.06504},
      archivePrefix={arXiv},
      primaryClass={stat.ME},
      url={https://arxiv.org/abs/2408.06504}, 
}

@misc{talts_sbc,
  
  url = {http://www.stat.columbia.edu/~gelman/research/unpublished/sbc.pdf},
  
  author = {Talts, Sean and Betancourt, Michael and Simpson, Daniel and Vehtari, Aki and Gelman, Andrew},
  
  title = {Validating Bayesian inference algorithms with simulation-based calibration},
  
  year = {2020},
}

@inproceedings{yao_domke_classifier_sbc_2023,
 author = {Yao, Yuling and Domke, Justin},
 booktitle = {Advances in Neural Information Processing Systems},
 editor = {A. Oh and T. Naumann and A. Globerson and K. Saenko and M. Hardt and S. Levine},
 pages = {36106--36131},
 publisher = {Curran Associates, Inc.},
 title = {Discriminative Calibration: Check Bayesian Computation from Simulations and Flexible Classifier},
 url = {https://proceedings.neurips.cc/paper_files/paper/2023/file/7103cd82de95a7b30983fcf74ba499ac-Paper-Conference.pdf},
 volume = {36},
 year = {2023}
}

@article{hem_makemyprior_2022,
 title={makemyprior: Intuitive Construction of Joint Priors for Variance Parameters in R},
 volume={110},
 url={https://www.jstatsoft.org/index.php/jss/article/view/v110i03},
 doi={10.18637/jss.v110.i03},
 number={3},
 journal={Journal of Statistical Software},
 author={Hem, Ingeborg and Fuglstad, Geir-Arne and Riebler, Andrea},
 year={2024},
 pages={1–39}
}

@misc{sailynoja2025posteriorsbc,
      title={Posterior SBC: Simulation-Based Calibration Checking Conditional on Data}, 
      author={Teemu Säilynoja and Marvin Schmitt and Paul-Christian Bürkner and Aki Vehtari},
      year={2025},
      eprint={2502.03279},
      archivePrefix={arXiv},
      primaryClass={stat.ME},
      url={https://arxiv.org/abs/2502.03279}, 
}

@article{geweke_getting_2004,
	title = {Getting it right},
	volume = {99},
	pages = {799--804},
	journal = {Journal of the American Statistical Association},
	year={2004},
	author = {Geweke, John}
}

@article{yu_assessment_2021,
	title = {Assessment and adjustment of approximate inference algorithms using the law of total variance},
	volume = {30},
	url = {https://doi.org/10.1080/10618600.2021.1880921},
	pages = {977--990},
	journal = {Journal of Computational and Graphical Statistics},
	author = {Yu, Xuejun and Nott, David J. and Tran, Minh-Ngoc and Klein, Nadja},
	year = {2021}
}

@article{gaffke2005,
  title={Three test statistics for a nonparametric one-sided hypothesis on the mean of a nonnegative variable},
  author={Gaffke, Norbert},
  journal={Mathematical Methods of Statistics},
  volume={14},
  number={4},
  pages={451--467},
  year={2005},
  publisher={Citeseer}
}

@misc{learnedmiller2020cimean,
      title={A New Confidence Interval for the Mean of a Bounded Random Variable}, 
      author={Erik Learned-Miller and Philip S. Thomas},
      year={2020},
      eprint={1905.06208},
      archivePrefix={arXiv},
      primaryClass={math.ST},
      url={https://arxiv.org/abs/1905.06208}, 
}

@article{wang2003noparametriclower,
title = {Nonparametric tests for the mean of a non-negative population},
journal = {Journal of Statistical Planning and Inference},
volume = {110},
number = {1},
pages = {75-96},
year = {2003},
issn = {0378-3758},
doi = {https://doi.org/10.1016/S0378-3758(01)00294-4},
url = {https://www.sciencedirect.com/science/article/pii/S0378375801002944},
author = {Weizhen Wang and Linda H. Zhao}
}

@article{sailynoja_graphical_2021,
author = {S\"{a}ilynoja, Teemu and B\"{u}rkner, Paul-Christian and Vehtari, Aki},
title = {Graphical Test for Discrete Uniformity and Its Applications in Goodness-of-Fit Evaluation and Multiple Sample Comparison},
year = {2022},
issue_date = {Apr 2022},
publisher = {Kluwer Academic Publishers},
address = {USA},
volume = {32},
number = {2},
issn = {0960-3174},
url = {https://doi.org/10.1007/s11222-022-10090-6},
doi = {10.1007/s11222-022-10090-6},
journal = {Statistics and Computing},
month = {apr},
numpages = {21}
}

@article{gelman2023recalibration,
  title={Understanding posterior recalibration for a simple example},
  author={Gelman, Andrew and Greengard, Philip and Gershunskaya, Julie and Savitsky, Terrance and Goodrich, Ben},
  year={2023},
  url = {https://sites.stat.columbia.edu/gelman/research/unpublished/calibration_toy.pdf},
  journal = "Unpublished manuscript."
}

@article{LIU200BFSensitivity,
title = {Bayes factors: Prior sensitivity and model generalizability},
journal = {Journal of Mathematical Psychology},
volume = {52},
number = {6},
pages = {362-375},
year = {2008},
issn = {0022-2496},
doi = {https://doi.org/10.1016/j.jmp.2008.03.002},
url = {https://www.sciencedirect.com/science/article/pii/S002224960800028X},
author = {Charles C. Liu and Murray Aitkin}
}

@article{dickey1970,
 ISSN = {00034851, 21688990},
 URL = {http://www.jstor.org/stable/2239734},
 author = {James M. Dickey and B. P. Lientz},
 journal = {The Annals of Mathematical Statistics},
 number = {1},
 pages = {214--226},
 publisher = {Institute of Mathematical Statistics},
 title = {The Weighted Likelihood Ratio, Sharp Hypotheses about Chances, the Order of a Markov Chain},
 urldate = {2025-06-04},
 volume = {41},
 year = {1970}
}

@Manual{bayesfactor_package,
    title = {BayesFactor: Computation of Bayes Factors for Common Designs},
    author = {Richard D. Morey and Jeffrey N. Rouder},
    year = {2024},
    note = {R package version 0.9.12-4.7},
    url = {https://CRAN.R-project.org/package=BayesFactor},
  }

@Article{bridgesampling_package,
    title = {{bridgesampling}: An {R} Package for Estimating Normalizing Constants},
    author = {Quentin F. Gronau and Henrik Singmann and Eric-Jan Wagenmakers},
    journal = {Journal of Statistical Software},
    year = {2020},
    volume = {92},
    number = {10},
    pages = {1--29},
    doi = {10.18637/jss.v092.i10},
  }

@article{
dimitraidis2021_reliability,
author = {Timo Dimitriadis  and Tilmann Gneiting  and Alexander I. Jordan },
title = {Stable reliability diagrams for probabilistic classifiers},
journal = {Proceedings of the National Academy of Sciences},
volume = {118},
number = {8},
pages = {e2016191118},
year = {2021},
doi = {10.1073/pnas.2016191118},
URL = {https://www.pnas.org/doi/abs/10.1073/pnas.2016191118},
eprint = {https://www.pnas.org/doi/pdf/10.1073/pnas.2016191118}}

@Article{Tsukamura2024_normalization,
author="Tsukamura, Yuki
and Okada, Kensuke",
title="The ``neglecting the vectorization`` error in Stan: erroneous coding practices for computing marginal likelihood and Bayes factors in models with vectorized truncated distributions",
journal="Behaviormetrika",
year="2024",
month="Jul",
day="01",
volume="51",
number="2",
pages="635--644",
issn="1349-6964",
doi="10.1007/s41237-024-00232-7",
url="https://doi.org/10.1007/s41237-024-00232-7"
}

@article{dimitriadis2023_calibration,
    author = {Dimitriadis, Timo and Dümbgen, Lutz and Henzi, Alexander and Puke, Marius and Ziegel, Johanna},
    title = {Honest calibration assessment for binary outcome predictions},
    journal = {Biometrika},
    volume = {110},
    number = {3},
    pages = {663-680},
    year = {2022},
    month = {12},
    issn = {1464-3510},
    doi = {10.1093/biomet/asac068},
    url = {https://doi.org/10.1093/biomet/asac068},
    eprint = {https://academic.oup.com/biomet/article-pdf/110/3/663/51111473/asac068\_supplementary\_data.pdf},
}

@article{ogle2020_identifiability_mixed,
author = {Ogle, Kiona and Barber, Jarrett J.},
title = {Ensuring identifiability in hierarchical mixed effects Bayesian models},
journal = {Ecological Applications},
volume = {30},
number = {7},
pages = {e02159},
doi = {https://doi.org/10.1002/eap.2159},
url = {https://esajournals.onlinelibrary.wiley.com/doi/abs/10.1002/eap.2159},
eprint = {https://esajournals.onlinelibrary.wiley.com/doi/pdf/10.1002/eap.2159},
year = {2020}
}

@article{brms,
 title={brms: An R Package for Bayesian Multilevel Models Using Stan},
 volume={80},
 url={https://www.jstatsoft.org/index.php/jss/article/view/v080i01},
 doi={10.18637/jss.v080.i01},
 number={1},
 journal={Journal of Statistical Software},
 author={Bürkner, Paul-Christian},
 year={2017},
 pages={1–28}
}

@article{stan,
 title={Stan: A Probabilistic Programming Language},
 volume={76},
 url={https://www.jstatsoft.org/index.php/jss/article/view/v076i01},
 doi={10.18637/jss.v076.i01},
 number={1},
 journal={Journal of Statistical Software},
 author={Carpenter, Bob and Gelman, Andrew and Hoffman, Matthew D. and Lee, Daniel and Goodrich, Ben and Betancourt, Michael and Brubaker, Marcus and Guo, Jiqiang and Li, Peter and Riddell, Allen},
 year={2017},
 pages={1–32}
}

@misc{SBC_package,
	title = {SBC: Simulation based calibration for rstan/cmdstanr models},
	url = {https://github.com/hyunjimoon/SBC/},
	author = {Kim, Shinyoung and Moon, Angie H. and Modrák, Martin and Säilynoja, Teemu},
	year = {2022}
}

@InProceedings{hartmann_prior_eliciation,
  title = 	 {Flexible Prior Elicitation via the Prior Predictive Distribution},
  author =       {Hartmann, Marcelo and Agiashvili, Georgi and B\"{u}rkner, Paul and Klami, Arto},
  booktitle = 	 {Proceedings of the 36th Conference on Uncertainty in Artificial Intelligence (UAI)},
  pages = 	 {1129--1138},
  year = 	 {2020},
  editor = 	 {Peters, Jonas and Sontag, David},
  volume = 	 {124},
  series = 	 {Proceedings of Machine Learning Research},
  month = 	 {03--06 Aug},
  publisher =    {PMLR},
  url = 	 {https://proceedings.mlr.press/v124/hartmann20a.html}
}

@inproceedings{key1999_model_choice,
    author = {Key, Jane T. and Pericchi, Luis R. and Smith, Adrian F. M.},
    title = {Bayesian model choice: What
and why.},
    booktitle = {Bayesian Statistics 6:Proceedings of the Sixth Valencia International Meeting},
    year = 1999,
pages ={343-370},
editor = {Bernardo, José M. and Berger, James O. and Dawid, A. P.  and Smith, Adrian F. M. }
}

@Article{Wagenmakers2018_bf_advantages,
author="Wagenmakers, Eric-Jan
and Marsman, Maarten
and Jamil, Tahira
and Ly, Alexander
and Verhagen, Josine
and Love, Jonathon
and Selker, Ravi
and Gronau, Quentin F.
and {\v{S}}m{\'i}ra, Martin
and Epskamp, Sacha
and Matzke, Dora
and Rouder, Jeffrey N.
and Morey, Richard D.",
title="Bayesian inference for psychology. Part I: Theoretical advantages and practical ramifications",
journal="Psychonomic Bulletin {\&} Review",
year="2018",
month="Feb",
day="01",
volume="25",
number="1",
pages="35--57",
issn="1531-5320",
doi="10.3758/s13423-017-1343-3",
url="https://doi.org/10.3758/s13423-017-1343-3"
}

@Article{Wagenmakers2023_jf_paradox,
author="Wagenmakers, Eric-Jan
and Ly, Alexander",
title="History and nature of the Jeffreys--Lindley paradox",
journal="Archive for History of Exact Sciences",
year="2023",
month="Jan",
day="01",
volume="77",
number="1",
pages="25--72",
issn="1432-0657",
doi="10.1007/s00407-022-00298-3",
url="https://doi.org/10.1007/s00407-022-00298-3"
}

@article{yao_stacking_2018,
author = {Yuling Yao and Aki Vehtari and Daniel Simpson and Andrew Gelman},
title = {{Using Stacking to Average Bayesian Predictive Distributions (with Discussion)}},
volume = {13},
journal = {Bayesian Analysis},
number = {3},
publisher = {International Society for Bayesian Analysis},
pages = {917 -- 1007},
year = {2018},
doi = {10.1214/17-BA1091},
URL = {https://doi.org/10.1214/17-BA1091}
}

@article{berger_pericchi_intrinsinc_bf_1996,
 ISSN = {01621459},
 URL = {http://www.jstor.org/stable/2291387},
 author = {James O. Berger and Luis R. Pericchi},
 journal = {Journal of the American Statistical Association},
 number = {433},
 pages = {109--122},
 publisher = {[American Statistical Association, Taylor & Francis, Ltd.]},
 title = {The Intrinsic Bayes Factor for Model Selection and Prediction},
 urldate = {2026-02-11},
 volume = {91},
 year = {1996}
}

@article{han2001bfmcmc,
 ISSN = {01621459, 1537274X},
 URL = {http://www.jstor.org/stable/2670258},
 author = {Cong Han and Bradley P. Carlin},
 journal = {Journal of the American Statistical Association},
 number = {455},
 pages = {1122--1132},
 publisher = {[American Statistical Association, Taylor & Francis, Ltd.]},
 title = {Markov Chain Monte Carlo Methods for Computing Bayes Factors: A Comparative Review},
 urldate = {2026-03-13},
 volume = {96},
 year = {2001}
}

@article{raftery1995bayesfactors,
author = {Robert E. Kass and Adrian E. Raftery},
title = {Bayes Factors},
journal = {Journal of the American Statistical Association},
volume = {90},
number = {430},
pages = {773--795},
year = {1995},
publisher = {Taylor \& Francis},
doi = {10.1080/01621459.1995.10476572}
}

@article{llorente2023bfreview,
author = {Llorente, F. and Martino, L. and Delgado, D. and L\'{o}pez-Santiago, J.},
title = {Marginal Likelihood Computation for Model Selection and Hypothesis Testing: An Extensive Review},
journal = {SIAM Review},
volume = {65},
number = {1},
pages = {3-58},
year = {2023},
doi = {10.1137/20M1310849}
}

@misc{mcewen2023machinelearningassistedbayesian,
      title={Machine learning assisted Bayesian model comparison: learnt harmonic mean estimator}, 
      author={Jason D. McEwen and Christopher G. R. Wallis and Matthew A. Price and Alessio Spurio Mancini},
      year={2023},
      eprint={2111.12720},
      archivePrefix={arXiv},
      primaryClass={stat.ME},
      url={https://arxiv.org/abs/2111.12720}, 
}

@article{spuriomancini2023bfsbi,
    author = {Spurio Mancini, A and Docherty, M M and Price, M A and McEwen, J D},
    title = {Bayesian model comparison for simulation-based inference},
    journal = {RAS Techniques and Instruments},
    volume = {2},
    number = {1},
    pages = {710-722},
    year = {2023},
    month = {11},
    issn = {2752-8200},
    doi = {10.1093/rasti/rzad051}    
}

@article{polanska2025bfsbi,
	author = {Polanska, Alicja and Price, Matthew A. and Piras, Davide and Spurio Mancini, Alessio and McEwen, Jason D.},
	journal = {The Open Journal of Astrophysics},
	doi = {10.33232/001c.146026},
	year = {2025},
	month = {oct 17},
	publisher = {Maynooth Academic Publishing},
	title = {Learned harmonic mean estimation of the {Bayesian} evidence with normalizing flows},
	volume = {8},
}

@article{Srinivasan2024bfsbi,
  title = {Bayesian evidence estimation from posterior samples with normalizing flows},
  author = {Srinivasan, Rahul and Crisostomi, Marco and Trotta, Roberto and Barausse, Enrico and Breschi, Matteo},
  journal = {Phys. Rev. D},
  volume = {110},
  issue = {12},
  pages = {123007},
  numpages = {14},
  year = {2024},
  month = {Dec},
  publisher = {American Physical Society},
  doi = {10.1103/PhysRevD.110.123007},
  url = {https://link.aps.org/doi/10.1103/PhysRevD.110.123007}
}

@inproceedings{plummer2003jags,
  title={JAGS: A program for analysis of Bayesian graphical models using Gibbs sampling},
  author={Plummer, Martyn},
  booktitle={Proceedings of the 3rd international workshop on distributed statistical computing},
  year={2003},
  organization={Vienna, Austria}
}

\section*{Appendix A - Theory and proofs}\label{proofs-and-theory}

We will denote the integral of \(f(x)\) w.r.t.~\(x\) over domain \(X\)
as \(\myint[_X]{x}{f(x)}\). The integral is implicitly over the relevant measure (i.e. counting measure for point masses and Borel measure for continuous distributions). \(\mathbb{I}[P]\) is the indicator
function for a given predicate \(P\). When a function can be understood
as describing a conditional probability distribution, we will use
\(\mid\) to separate the function arguments we condition on. This is
only to assist comprehension and has the same semantic meaning as using
a comma.

All definitions and proofs implicitly assume a
single model \(\pi\), data space \(Y\) and parameter
space \(\Theta\) are given. We use $\pi$ to indicate various densities w.r.t. the model, once again implicitly assuming a relevant measure. Unless otherwise noted, expectations and probabilities are taken against the joint distribution over $\theta \times Y$ implied by $\pi$.

\begin{definition}[Posterior family] Given a data space $Y$ and a parameter space $\Theta$, a \emph{posterior family} $\phi$ assigns a normalized posterior density to each possible $y \in Y$. I.e. posterior family is a function $\phi : \Theta \times Y  \rightarrow  \mathbb{R^{+}}$ such that
\begin{equation}
\forall y: \myint[_\Theta]{\theta}{\phi(\theta \mid y)} = 1.
\end{equation}
We will also use $\phi(y)$ to denote the implied distribution over $\Theta$ given a specific $y \in Y$.
\end{definition}

\begin{definition}[Test quantity] A \emph{test quantity} is any measurable function $f : \Theta \times Y \to \mathbb{R}$. A \emph{binary test quantity} is the special case when $f : \Theta \times Y \to \{0,1\}$. We will use $f$ to also denote the implied random variable over $\Theta \times Y$.
\end{definition}

\begin{definition}[binary prediction calibration]  Given posterior family \(\phi\) and binary test quantity $f$, the \emph{binary prediction} $D_{\phi,f}(y) = \mathbb{E}_{\phi(y)}[f(\theta, y) \mid y] = \myint[_\Theta]{\theta}{\phi(\theta \mid y) \mathbb{I}\left[ f(\theta, y) = 1\right]}$. We will use $D_{\phi,f}$ to denote the implied random variable over $\Theta \times Y$. When there is no risk of confusion we will omit the lower indices and just use $D$.

A posterior family \(\phi\) \emph{satisfies binary prediction
calibration check} w.r.t. $f$  if

\begin{equation}
\forall d \in [0,1] : \mathbb{E}[f\mid D_{\phi,f} = d] = d
\end{equation}

wherever the left-hand side is well-defined (i.e.~when there exist $y \in Y$ such that $\pi_\text{marg}(y) > 0$ and $D_{\phi,f}(y) = d$). It is equivalent to $\mathbb{E}[f\mid D_{\phi,f}] = D_{\phi,f}$ $\pi$-almost surely.
\end{definition}

\begin{definition}[CDF, tie probability, quantile functions]
Following \citet{modrak2025}, we first define
\begin{itemize}
    \item fitted CDF: $C_{\phi,f}: \bar{\mathbb{R}}\times Y\to [0,1]$, \\ 
    $C_{\phi,f}(s \mid y) := \mbox{Pr}_{\phi(y)}(f \leq s \mid y) = \myint[_{\Theta}]{\theta}{\,\mathbb{I}\left[f\left(\theta, y \right) \leq s\right]\phi\left(\theta\mid y\right)}$
    \item fitted tie probability: $D_{\phi,f}: \bar{\mathbb{R}}\times Y\to [0,1]$, \\
    $D_{\phi,f}(s \mid y) := \mbox{Pr}_{\phi(y)}(f = s \mid y) = \myint[_\Theta]{\theta}{\phi(\theta \mid y) \mathbb{I}\left[ f(\theta, y) = s \right]}$ 
\end{itemize}
Note that the definition is consistent with that of binary prediction, specifically for $f$ binary we have $D_{\phi, f}(1 \mid y) = D_{\phi,f}(y)$.
\end{definition}

\begin{definition}[Continuous rank CDF, continuous $q$, continuous SBC]
Assume a data space $Y$, a parameter space $\Theta$, a test quantity $f$, and a posterior family $\phi$. For fixed $\tilde\theta \in \Theta$ and $y \in Y$ we define the \emph{continuous rank CDF} $r_{\phi,f} : [0,1] \times \Theta \times Y \to [0, 1]$ as
\begin{equation}
r_{\phi,f}(x \mid \tilde\theta, y) := 
\mbox{Pr}_U\left(C_{\phi,f} \left(f \left(\left.\tilde\theta, y \right) \:\right|\: y \right) - U D_{\phi,f} \left(f \left(\left.\tilde\theta, y \right) \:\right|\: y \right) \leq x\right) ,   
\end{equation}
assuming $U$ is a random variable distributed uniformly over the $[0, 1]$ interval. If the fitted tie probability is $0$, then $r$ is a step function and the implied rank distribution is degenerate.

We then define the \emph{continuous } $q: [0,1] \times Y \to [0,1]$ as
\begin{equation}
    q_{\phi,f}(x \mid y) := 
    \mathbb{E}[r_{\phi,f}(x\mid y)]
= \myint[_\Theta]{\tilde\theta}{\pi_\text{post}(\tilde\theta \mid y) r_{\phi,f}(x \mid \tilde\theta, y)},
\end{equation}

Finally, \emph{$\phi$ passes continuous SBC w.r.t.\ $f$} if $
\forall x \in [0, 1]: \mathbb{E}[q_{\phi,f}(x)] = \myint[_Y]{y}{q_{\phi,f}(x\mid y) \pi_\text{marg}(y)}  = x$.
\end{definition}

The above definition is a bit tedious, but a quick intuition is that $q$ is the quantile-quantile function of ranks of the true values within the candidate posterior implied by $\phi$ with linear smoothing over ties.

\citet{modrak2025} also define \emph{sample SBC} where $\phi$ and $\pi$ are only accessible via draws from the relevant distributions, which is how SBC is implemented in practice. Here, we will work only with continuous SBC, which is closely related to sample SBC (specifically, passing continuous SBC is equivalent to passing sample SBC for any number of draws taken). Continuous SBC is also more relevant for Bayes factors as we typically obtain (an approximation of) the
posterior model probability (i.e. $D(y)$ in our notation), not just draws of the model index.

\begin{theorem} For any binary test quantity $f$, if posterior family \(\phi\)
 satisfies binary prediction
calibration check w.r.t \(f\) it also passes continuous SBC w.r.t $f$. If the distribution of binary predictions $D_{\phi,f}$ is not singular and $\phi$ passes continuous SBC w.r.t $f$, $\phi$ also satisfies binary prediction
calibration check w.r.t $f$.
\end{theorem}

\begin{proof} 
For a uniformly distributed random variable \(U \sim \mathcal{U}(0,1)\),
it holds $c-Ud \sim \mathcal{U}(c-d,c)$ for any $d>0$ and any $c\in\mathbb{R}$, whereas the random variable $c-Ud$ has the cumulative distribution function
\begin{align}
    \mbox{Pr}_U(c-U d \le x) =\frac{x-(c-d)}{d}\mathbb{I}\left[c-d\le x < c\right]+\mathbb{I}\left[c\le x\right]
\end{align}
for any $x\in\mathbb{R}$. Since the function $f$
only takes values in $\{0,1\}$, 
it holds $f=\mathbb{I}[f=1]$ and
\begin{equation}
\begin{aligned}
    D_{\phi,f}(f(\theta,y)\mid y) 
    &= f(\theta,y)D_{\phi,f}(1\mid y)+(1-f(\theta,y))(1-D_{\phi,f}(1\mid y)),\\
    C_{\phi,f}(f(\theta,y)\mid y) - D_{\phi,f}(f(\theta,y)\mid y) 
    &= f(\theta,y)(1-D_{\phi,f}(1\mid y)),\\
    C_{\phi,f}(f(\theta,y)\mid y) 
    &= f(\theta,y) + (1-f(\theta,y))(1-D_{\phi,f}(1\mid y))
\end{aligned}
\end{equation}
for any $\theta\in\Theta$ and any $y\in Y$. Thus, according to the above representation of the cumulative distribution function of $c-Ud$, it holds
\small
\begin{equation}\begin{aligned}
    r_{\phi,f}(x\mid \theta,y) 
    &= \mbox{Pr}_U(C_{\phi,f}(f(\theta,y)\mid y)-U D_{\phi,f}(f(\theta,y)\mid y) \le x)\\
    &= f(\theta,y)\left(\frac{x-(1-D(\theta, y))}{D(\theta, y)}\mathbb{I}\left[1-D(\theta, y)\le x < 1\right] + \mathbb{I}\left[1\le x\right]\right)\\
    &\quad + (1-f(\theta,y))\left(\frac{x}{1-D(\theta,y)}\mathbb{I}\left[0\le x < 1-D(\theta, y)\right]+\mathbb{I}\left[1-D(\theta, y)\le x\right]\right)
\end{aligned}\end{equation}
\normalsize
for any $x\in\mathbb{R}$, where $D(\theta, y)=D_{\phi,f}(1\mid y)$ as mentioned in the definition of binary prediction calibration. We note that $r_{\phi,f}(0  \mid \theta ,y) = 0$ and $r_{\phi,f}(1 \mid \theta ,y) = 1$. 
By taking the conditional expectation $\mathbb{E}[\ \cdot\mid D]$ with respect to the probability measure on the Bayesian model $\pi$, you find
\begin{align}
    \mathbb{E}[ r_{\phi,f}(x)  \mid D]
    &= \mathbb{E}[f\mid D]\left(\frac{x-(1-D)}{D}\mathbb{I}\left[1-D\le x < 1\right] + \mathbb{I}\left[1 \leq x\right]\right) \nonumber \\
    &\quad + (1-\mathbb{E}[f\mid D])\left(\frac{x}{1-D}\mathbb{I}\left[0\le x < 1-D\right]+\mathbb{I}\left[1-D \le x\right] \right)
    \label{eq:expected_rank}
\end{align}
for any $x\in\mathbb{R}$. 

For the first implication, assume that $\phi$ satisfies the binary prediction calibration check with respect to $f$, it holds
    $\mathbb{E}[f\mid D] 
    = D$
and consequently 
\begin{equation}
    \mathbb{E}[ r_{\phi,f}(x)  \mid D] = x \mathbb{I}\left[0\le x \le 1\right]+ \mathbb{I}\left[1<x\right]
\end{equation}
for any $x\in\mathbb{R}$. Therefore
\begin{equation}
    \mathbb{E}[r_{\phi,f}(x)] = \mathbb{E}[\mathbb{E}[r_{\phi,f}(x) \mid D]] = x
\end{equation}
for any $x\in[0,1]$, meaning that $\phi$ passes continuous SBC with respect to $f$.

For the second implication, we assume the distribution of $D_{\phi,f}$ is not singular and thus can be described by a mixture of a probability distribution function and a probability mass function. This also implies that the CDF of the distribution of $D_{\phi,f}$ is equal to the derivative of its integral almost everywhere.
We note that $r_{\phi,f}(x)=r_{\phi,f}(x\mid \theta, y)$ only depends on $\theta, y$ via $ f = f(\theta, y)$ and $  D = D(\theta,y)$. So, we have 
%
\begin{equation}
\begin{aligned}
    r_{\phi,f}(x)
    &=f\left(\frac{x-(1-D)}{D}
    \mathbb{I}\left[1-D \le x < 1\right]
    + 
    \mathbb{I}\left[1 \le x \right]\right)
    \\
    &\quad +  (1-f)\left(\frac{x}{1-D}\mathbb{I}\left[0\le x < 1-D\right]
    +\mathbb{I}\left[1-D\le x \right] \right)
\end{aligned}
\end{equation}
We denote $m = \mbox{Pr}(f = 1)$, \(G_0(x) = \mbox{Pr} (D \leq x \mid  f = 0)\) and \(G_1(x) = \mbox{Pr}(D \leq x \mid  f = 1)\).
For $s<0$ you find $\mbox{Pr}(r_{\phi,f}(x) \leq s) = 0$ and for $s\ge 1$ it holds $\mbox{Pr}(r_{\phi,f}(x) \leq s) = 1$. For the remaining $s\in[0,1)$ note that $\{x<1-D,f =1\}\subset\{(x-(1-D))/D \le s,f=1\}$ 
and thus
\small
\begin{equation}
\begin{aligned}
    F^{r_{\phi,f}(x)}(s) :=\mbox{Pr}(r_{\phi,f}(x) \leq s)
    &= \mbox{Pr}\left( \frac{x-(1-D)}{D}\le s, f = 1\right) + \mbox{Pr}\left(\frac{x}{1-D}\le s, f = 0\right)\\
    &= \mbox{Pr}\left( D\le \frac{1-x}{1-s}, f = 1\right) + \mbox{Pr}\left( D\le \frac{s-x}{s}, f = 0\right) \\
    &= m G_1\left(\frac{1 - x}{1 - s}\right) 
    + (1 - m)G_0 \left(\frac{s - x}{s}\right) 
\end{aligned}
\end{equation}
\normalsize
Recall that $D\in[0,1]$, such that $G_0(s)=G_1(s)=0$ for $s<0$ and $G_0(s)=G_1(s)=1$ for $s\ge 1$. Thus, you get the representation
\small
\begin{equation}
    F^{r_{\phi,f}(x)}(s)
    = m G_1\left(\frac{1 - x}{1 - s}\right)\mathbb{I}\left[0\le s < x\right] + \left(m + (1 - m)G_0 \left(\frac{s - x}{s}\right)\right)\mathbb{I}\left[x\le s < 1\right] + \mathbb{I}\left[1 \le s\right]
\end{equation}
\normalsize
which yields
\begin{equation}
\begin{aligned}
    1-F^{r_{\phi,f}(x)}(s)
    = \mathbb{I}\left[s < 0\right] & + \left(1-m G_1\left(\frac{1 - x}{1 - s}\right)\right)\mathbb{I}\left[0\le s < x\right]\\
    & + (1- m) \left(1 - G_0 \left(\frac{s - x}{s}\right)\right)\mathbb{I}\left[x\le s < 1\right].
\end{aligned}
\end{equation}
%
%
For a non-negative random variable $X$, it holds
\begin{equation}\begin{aligned}
    \mathbb{E}[X] =\mathbb{E}\left[\myint[_0^X]{s}{1} \right]  =  =\mathbb{E}\left[\myint[_0^\infty]{s}{\mathbb{I}\left[X>s\right]} \right] =\myint[_0^\infty]{s}{\mathbb{P}(X>s)} = \myint[_0^\infty]{s}{(1-F^X(s))}
\end{aligned}\end{equation}
by Tonelli's theorem, such that when $\phi$ passes continuous SBC with respect to $f$, you find
%
\begin{equation}\begin{aligned}
    x = \mathbb{E}[r_{\phi,f}(x)]
    &= \myint[_0^\infty]{s}{(1-F^{r_{\phi,f}(x)}(s))} \\
    &= \myint[_0^x]{s}{\left(1-m G_1\left(\frac{1 - x}{1 - s}\right)\right)}
    +\myint[_x^1]{s}{(1- m) \left(1 - G_0 \left(\frac{s - x}{s}\right)\right)}\\
    &= x -m \myint[_0^x]{s}{G_1\left(\frac{1 - x}{1 - s}\right)}+(1 - m) \myint[_x^1]{s}{ \left(1 - G_0 \left(\frac{s - x}{s}\right)\right)},
\end{aligned}\end{equation}
and thus
\begin{align}
    \myint[_x^1]{s}{\left(1- G_0 \left(\frac{s - x}{s}\right)\right)} = \frac{m}{1 - m} \myint[_0^x]{s} {G_1\left(\frac{1-x}{1 -s}\right)}.
    \label{eq:inteqwithm}
\end{align}
%
Recall that for a differentiable $h:[a,b]\rightarrow\mathbb{R}$ and a function $F$ which is not singular and has an antiderivative almost everywhere
\begin{equation}\begin{aligned}
    \myint[_{h(a)}^{h(b)}]{s}{F(s)}=\myint[_a^b]{s}{F(h(s))h'(s)}.
\end{aligned}\end{equation}
For the left-hand side of \eqref{eq:inteqwithm} we choose $h_0(s)=\frac{x}{1-s}$ so that we have $h_0\left(\frac{s-x}{s}\right) = s$. 
Integration by substitution with $h_1$ yields 
\begin{equation}\begin{aligned}
\myint[_x^1]{s}{\left(1-G_0\left( \frac{s-x}{s}\right)\right)}
&=\myint[_0^{1-x}]{s}{\left( 1-G_0(s) \right)\frac{x}{(1-s)^2}}\\
&=x\myint[_0^{1-x}]{s}{\left( 1-G_0(s) \right)\frac{1}{(1-s)^2}}.
\end{aligned}\end{equation}
Then, for the right-hand side of \eqref{eq:inteqwithm} we choose $h_1(s) = 1-\frac{1-x}{s}$ so that $h_1\left(\frac{1-x}{1-s}\right) = s$.
Integration by substitution with $h_1$ and then with $h_2(s) =  1-s$ yields
\begin{equation}\begin{aligned}
\myint[_0^x]{s}{G_1\left(\frac{1-x}{1-s}\right)}
=\myint[_{1-x}^1]{s}{G_1\left(s\right)\frac{1-x}{s^2}}
&=(1-x)\myint[_{1-x}^1]{s}{G_1\left(s\right)\frac{1}{s^2}}\\
&=(1-x)\myint[_{0}^x]{s}{G_1(1-s)\frac{1}{(1-s)^2}}.
\end{aligned}\end{equation}
Therefore
\begin{equation}
\myint[_0^{1-x}]{s}{\left( 1-G_0(s) \right)\frac{1}{(1-s)^2}}=\frac{m}{1 - m}\frac{1-x}{x}\myint[_0^x]{s}{G_1\left(1-s\right)\frac{1}{(1-s)^2}}
\end{equation}
By the fundamental theorem of calculus, recall that for any function $h$ continuous almost everywhere, but not singular, the derivative of $H(x)=\myint[_0^x]{s}{h(s)}$ is almost everywhere given by $H'(x) = h(x)$ 
and the derivative of $g(x)=1-x$ is given by $g'(x) = -1$. By the chain rule, the function $H(g(x))=\myint[_0^{1-x}]{s}{h(s)}$ has the derivative $H'(g(x))g'(x)=-h(1-x)$.
Differentiating both sides of the above equation with respect to $x$ results in
\begin{equation}
\left(G_0(1-x)-1\right)\frac{1}{x^2}=\frac{m}{1 - m}\left(\frac{1-x}{x}G_1(1-x)\frac{1}{(1-x)^2}-\frac{1}{x^2}\myint[_{0}^x]{s}{G_1(1-s)\frac{1}{(1-s)^2}}\right)    
\end{equation}

and
\begin{equation}
G_0(1-x)=1+\frac{m}{1 - m}\left(\frac{x}{1-x}G_1(1-x)-\myint[_0^x]{s}{G_1(1-s)\frac{1}{(1-s)^2}}\right)
\end{equation}
almost everywhere. Consequently,
\begin{equation}
    G_0(x)=G_0(1-(1-x))=1+\frac{m}{ 1 - m}\left(\frac{1-x}{x}G_1(x)-\myint[_0^{1-x}]{s}{G_1(1-s)\frac{1}{(1-s)^2}}\right)
\end{equation}
almost everywhere. Since two CDFs that are equal almost everywhere need to be equal everywhere, the above needs to actually hold for all $x \in [0,1]$.
We note that the right-hand side of the above equation is differentiable at $x$ if and only if $G_1$ is differentiable at $x$. Therefore for the equality to hold for all $0 < x < 1$, $G_0(x)$ has to be differentiable exactly at those $x$ where $G_1(x)$ is differentiable. The same holds for continuity. If $G_0, G_1$ are differentiable at $x$, we differentiate again w.r.t $x$ and obtain
\begin{equation}\begin{aligned}
g_0(x) &=\frac{m}{1 - m}\left(\frac{1-x}{x}g_1(x)-\frac{1}{x^2}G_1(x)+\frac{1}{x^2}G_1(x)\right)
=\frac{m}{1 - m}\frac{1-x}{x}g_1(x) 
\end{aligned}\end{equation}

Since $G_i$ have to be differentiable almost everywhere, and two probability density functions that differ only on a set of measure zero will describe the same continuous distribution, we thus know the ratio of the two densities wherever the distribution can be described by a density.

When $G_0$ and $G_1$ are discontinuous at $x$, there is a point mass (in both distributions) at $x$ and we have $\mbox{Pr}(D = x \mid f = i) = 
G_i(x) - \lim_{s \to x-} G_i(s)$. The integral $\myint[_0^x]{s}{G_1(1-s)\frac{1}{(1-s)^2}}$ is continuous in $x$ and thus does not contribute to the difference of the one-sided limits and we have a result analogous to the one for densities:
\begin{equation}\begin{aligned}
    \mbox{Pr}(D = x \mid f = 0) &= 
    G_0(x)- \lim_{s \to x-} G_0(s) \\
    &= \frac{m}{1 - m}\frac{1-x}{x}\left(
    G_1(x)- \lim_{s \to x-} G_1(s)\right)\\
    &= \frac{m}{1 - m}\frac{1-x}{x} \mbox{Pr}(D = x \mid f = 1)
\end{aligned}\end{equation}
This means that for the possibly mixed discrete-continuous conditional density of $D$ we obtain
\begin{equation}
    \pi(D = d \mid f = 0) = \frac{(1 - d)\mbox{Pr}(f = 1)}{d\mbox{Pr}(f = 0)} \pi(D = d \mid f = 1)
\end{equation}
and by Bayes theorem we have:
\begin{align}
\mbox{Pr}(f = 1 \mid D = d) &= \frac{\pi(D = d \mid f = 1) \mbox{Pr}(f = 1)}{
\pi(D = d \mid f = 0) \mbox{Pr}(f = 0) + \pi(D = d \mid f = 1) \mbox{Pr}(f = 1)} \notag \\
&=\frac{\pi(D = d \mid f = 1) \mbox{Pr}(f = 1)}{
\frac{(1 - d)\mbox{Pr}(f = 1)}{d\mbox{Pr}(f = 0)} \pi(D = d \mid f = 1) \mbox{Pr}(f = 0) + \pi(D = d \mid f = 1) \mbox{Pr}(f = 1)} \notag  \\
&= \frac{1}{\frac{1-d}{d} + 1} = d
\end{align}
And thus binary calibration holds.

\end{proof}

\begin{remark}
This equivalence between SBC and binary prediction calibration implies that all limitations of SBC directly apply to
binary calibration. Notably, following Theorem 7 of \cite{modrak2025}, if
\(f\) ignores data (e.g.~when we do binary calibration for the model
index), it will be satisfied when \(\phi = \pi_{\text{prior}}\). More generally, whenever the value of $f$ does not change when we modify some aspect of the data, a model that treats this aspect as unobserved will pass SBC and satisfy the binary prediction calibration check w.r.t $f$.
\end{remark}

\begin{definition}[Data-averaged posterior] A posterior family $\phi$ has the correct data-averaged posterior for a test quantity $f$ if for all $s \in \mathbb{R}$
\begin{equation}
    \forall s\in\mathbb{R}: \mbox{Pr}(f \leq s) = \mathbb{E}[\mbox{Pr}_{\phi}(f \leq s\mid y)]
\end{equation}
\end{definition}
The definition is equivalent to the explicit integral formula in \citet{modrak2025} stating
\begin{gather}
\myint[_Y]{y}{ \myint[_\Theta]{\theta}{\mathbb{I}\left[ f(\theta, y) < s \right] \pi_\text{obs}(y\mid\theta) \pi_\text{prior}(\theta)}} \notag \\
= \myint[_\Theta]{\theta}{ \myint[_Y]{y}{ \myint[_\Theta]{\tilde\theta}{ \mathbb{I}\left[f(\theta, y) < s\right] \phi(\theta \mid y) \pi_\text{obs}(y\mid\tilde\theta) \pi_\text{prior}(\tilde\theta)}}}.
\end{gather}

\begin{theorem}
For any binary test quantity $f$ if posterior family \(\phi\) satisfies binary prediction
calibration check w.r.t \(f\) it will have the correct data-averaged posterior for $f$.
\end{theorem}

\begin{proof}

We note that for binary test quantities the CDF of both prior and the data-averaged posterior is defined by a single number:
\begin{equation}\begin{aligned}
   &\forall s\in\mathbb{R}: \mbox{Pr}(f \leq s) = \mathbb{E}[\mbox{Pr}_{\phi}(f \leq s\mid y)] \\
   &\quad \iff \mbox{Pr}(f = 1) = \mathbb{E}[\mbox{Pr}_{\phi}(f = 1\mid y)]
\end{aligned}\end{equation}
Since $\phi$ satisfies the binary prediction calibration check, it holds $\mathbb{E}[f\mid D_{\phi,f}] = D_{\phi,f}$ and thus
\begin{equation}\begin{aligned}
\mbox{Pr}(f = 1) &= \mathbb{E}[f] = \mathbb{E}[\mathbb{E}[f\mid D_{\phi,f}]] = \mathbb{E}[D_{\phi,f}] = \mathbb{E}[\mbox{Pr}_{\phi}(f = 1\mid y)]
\end{aligned}\end{equation}
\end{proof}

\begin{remark}[Rejecting datasets]
Assume that we have a function $\mathtt{accept}: Y \to [0,1]$ denoting the probability that a given data point $y$ is accepted.
We define a variable $A \sim \text{Bernoulli}(\mathtt{accept}(y))$. This implies a joint distribution $\pi_\text{joint}(\theta, y, a)$ that factorizes as $\pi_\text{joint}(\theta, y, a) = \pi(a \mid y)\pi_\text{obs}(y \mid \theta)\pi_\text{prior}(\theta)$. 
It is a well-known fact of Bayesian statistics that the posterior does not change at all when we 
condition on accepting a dataset:
\begin{gather}
\pi_\text{post}(\theta \mid y, A = 1) = \frac{\pi(A = 1 \mid y) \pi_\text{obs}(y \mid \theta)\pi_\text{prior}(\theta)}{\myint[_\Theta]{\tilde\theta}{\pi(A = 1 \mid y) \pi_\text{obs}(y \mid \tilde\theta)\pi_\text{prior}(\tilde\theta)}} = 
\frac{\pi_\text{obs}(y \mid \theta)\pi_\text{prior}(\theta)}{\myint[_\Theta]{\tilde\theta}{ \pi_\text{obs}(y \mid \tilde\theta)\pi_\text{prior}(\tilde\theta)}} \notag \\= \pi_\text{post}(\theta \mid y)
\end{gather}
It follows that there is no need to adjust the SBC or binary prediction calibration check when conditioning on $A$.
\end{remark}

\section*{Appendix B - Notes on test quantities}

The choice of test quantities beyond the model index  ($f_1, \dots, f_J$) in SBC (Algorithm 1) deserve further comment: the simplest test quantities to use are just individual elements of the parameter vector. For a BMA model, this notion is a bit problematic, as different submodels may have different parameters. When a parameter exists and plays the same role in both models, we have found it useful to mix the posterior draws from all models (the same way one would pool BMA predictions). When a parameter is missing from a model $i$, we have found it useful to treat any draw of that parameter under model $i$ as $-\infty$. This lets us include at least some check for all parameters in all models in the default SBC procedure and lets us assume the parameter vector is the same length in all simulations/posterior draws. 

As in standard SBC, adding suitable data-dependent quantities may substantially improve sensitivity. The likelihood of the data under the chosen model is a good choice if available (see \citeinparen{modrak2025}).

\section*{Appendix C - Metrics to assess data-averaged
posterior}\label{metrics-to-assess-data-averaged-posterior}
    
Here, we compare the performance of the test of \citet{gaffke2005} to
a standard one-sample t-test. To compare to the work of Schad et al., we also used the
Bayesian t-test, where we set the scale parameter for the alternative
hypothesis to $\frac{1}{12}$ -- the sd of uniform distribution over \([0,1]\) or to standard medium or wide prior ($\frac{\sqrt{2}}{2}$ and $\frac{3}{2}$ respectively). We treat Bayes factor $>10$ in favor of the alternative as evidence of miscalibration. 

When the prior probability of $\mathcal{M}_1$ is not known analytically (e.g., due to rejection sampling), we need to use a two-sample t-test against the draws from the prior (i.e. the simulated chosen model). Since the chosen model should typically correlate with its posterior probability a paired test is preferable. A paired version of the Gaffke is easy to construct by testing whether the mean of the difference is zero, with the difference bounded within $[-1, 1]$.

\begin{figure}

\centering{

\includegraphics[width=\textwidth]{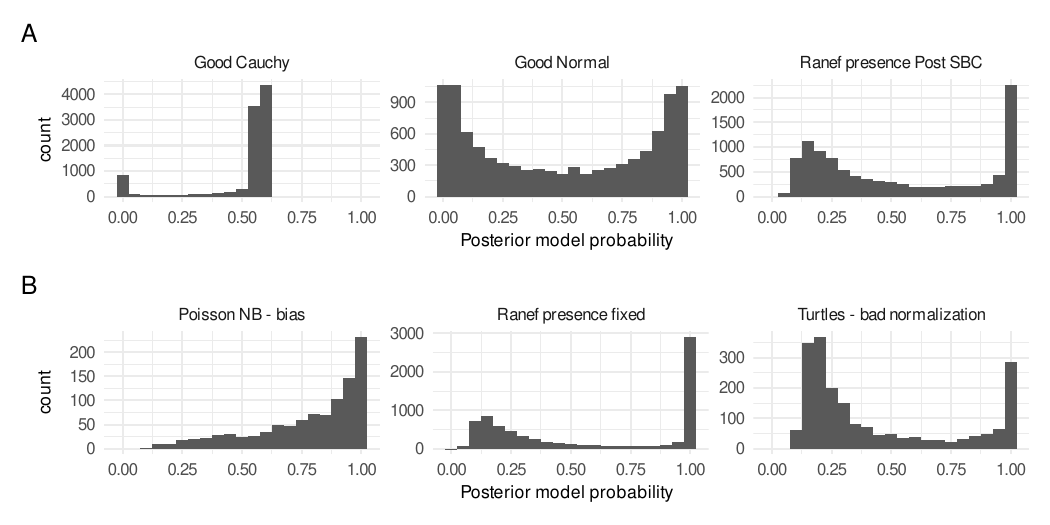}

}

\caption{\label{fig-dap-probs}Overall distribution of posterior model probabilities A) for scenarios where the model is calibrated, B) scenarios where DAP should detect a problem.}

\end{figure}

We examine all scenarios from the main part of the paper, where the Bayes factors are calibrated as well as those where there is a problem that is detectable by the date-averaged posterior checking. Figure~\ref{fig-dap-probs} shows the overall distribution of posterior model probabilities in all cases. Table~\ref{tab-dap-correct} shows the false positive rates when the model is in fact correct and Table~\ref{tab-dap-wrong} shows the power when the computation is wrong. We see that when the number of simulations is small, there are scenarios where the standard t-test has inflated false positive rates but with 100 simulations, standard t-test achieves nominal false positive rates in all scenarios. As expected, the Gaffke approach has proper error control in all situations, although it can be conservative. Across all prior scales, Bayesian t-test shows both inflated as well as overly conservative error rates. With regards to power, the paired variant is superior to unpaired. The standard t-test has the highest power in all situations, with the Gaffke approach catching up as the number of simulations increases, while Bayesian t-test has the lowest power in all scenarios except one (Turtles - bad normalization with N = 10) regardless of prior scale. The ordering of methods does not change if we instead use $3$ as a threshold for the Bayes factor.
    
\newpage
\begin{table}
\caption{\label{tab-dap-correct}False positive rates for various tests for data-averaged posterior in scenarios where the model is calibrated. All tests averaged over 1000 runs with the remaining uncertainty at most 2.7 percentage points.\\}  
\begin{tabular}{lrrrrrr}
\toprule
 &  &  &  & \multicolumn{3}{c}{Bayesian t-test} \\
Scenario & N & T-test & Gaffke & $r = \frac{1}{12}$ & $r = \frac{\sqrt{2}}{2}$ & $r = \frac{3}{2}$ \\
\midrule
Good Cauchy & 10 & 22.9\% & 0.0\% & 14.6\% & 18.1\% & 18.2\% \\
& 20 & 10.2\% & 0.1\% & 5.7\% & 6.3\% & 6.1\% \\
& 50 & 7.5\% & 1.3\% & 1.0\% & 1.5\% & 1.1\% \\
& 100 & 7.2\% & 2.3\% & 0.3\% & 0.3\% & 0.2\% \\
Good Normal & 10 & 4.9\% & 1.5\% & 0.5\% & 0.9\% & 0.9\% \\
& 20 & 5.3\% & 2.4\% & 0.1\% & 0.3\% & 0.3\% \\
& 50 & 5.0\% & 3.6\% & 0.0\% & 0.2\% & 0.0\% \\
& 100 & 5.2\% & 3.4\% & 0.2\% & 0.3\% & 0.1\% \\
Ranef presence Post SBC & 10 & 6.2\% & 0.9\% & 0.3\% & 0.8\% &
0.8\% \\
& 20 & 4.9\% & 2.2\% & 0.1\% & 0.3\% & 0.2\% \\
& 50 & 4.9\% & 3.1\% & 0.0\% & 0.1\% & 0.0\% \\
& 100 & 5.6\% & 3.1\% & 0.0\% & 0.0\% & 0.0\% \\
\bottomrule
\end{tabular}
\end{table}

\begin{table}
\caption{\label{tab-dap-wrong}Power for various tests for data-averaged posterior in scenarios where the model is not calibrated. All tests averaged over 1000 runs with the remaining uncertainty at most 3.1 percentage points. For the turtles scenario, the prior probability is not known analytically due to rejection sampling and thus a two sample test is performed, either paired (the default) or unpaired. There is no unpaired two sample version of the Gaffke test.\\}   
\centering
\begin{tabular}{lrrrrrr}
\toprule
 &  &  &  & \multicolumn{3}{c}{Bayesian t-test} \\
Scenario & N & T-test & Gaffke & $r = \frac{1}{12}$ & $r = \frac{\sqrt{2}}{2}$ & $r = \frac{3}{2}$ \\
\midrule
Poisson NB - bias & 10 & 88.2\% & 59.5\% & 22.5\% & 44.7\% &
45.2\% \\
& 20 & 99.4\% & 98.7\% & 79.9\% & 93.2\% & 91.5\% \\
& 50 & 100.0\% & 100.0\% & 100.0\% & 100.0\% & 100.0\% \\
& 100 & 100.0\% & 100.0\% & 100.0\% & 100.0\% & 100.0\% \\
Ranef presence fixed & 10 & 9.9\% & 3.4\% & 0.4\% & 1.2\% &
1.2\% \\
& 20 & 16.4\% & 10.8\% & 0.4\% & 1.9\% & 1.3\% \\
& 50 & 34.3\% & 28.5\% & 2.8\% & 5.1\% & 3.3\% \\
& 100 & 60.6\% & 56.3\% & 12.3\% & 14.2\% & 9.8\% \\
Turtles & 10 & 9.0\% & 0.1\% & 0.3\% & 1.0\% & 1.0\%\\
\ \  - bad normalization & 20 & 18.0\% & 4.1\% & 0.2\% & 1.3\% & 0.9\%\\
\ \ (paired) & 50 & 41.2\% & 28.3\% & 6.6\% & 9.9\% & 7.3\%\\
 & 100 & 75.5\% & 66.9\% & 29.6\% & 30.0\% & 22.4\%\\
Turtles& 10 & 3.7\% & - & 0.1\% & 0.8\% & 0.8\%\\
\ \  - bad normalization & 20 & 5.7\% & - & 0.2\% & 0.6\% & 0.6\%\\
\ \ (unpaired) & 50 & 17.3\% & - & 0.6\% & 1.3\% & 0.9\%\\
 & 100 & 44.4\% & - & 2.9\% & 6.0\% & 3.0\%\\
 \bottomrule
\end{tabular}
\end{table}

\clearpage 

\section*{Appendix D - Good check convergence}\label{good-check-convergence}

For example, assume the following models with no parameters and a single
observation \(y \in \mathbb{R}\):
\begin{equation}
\begin{split}
\mathcal{M}_0: y \sim \text{Cauchy}(0, 1) & \hspace{4em}
\mathcal{M}_1: y \sim N(0, 1) 
\end{split}
\end{equation}
we have

\begin{equation}
BF_{0,1}(y) = \frac{\text{Cauchy}(y \mid  0, 1)}{\text{Norm}(y \mid  0,1)} = \sqrt{\frac{2}{\pi}} \frac{e^\frac{y^2}{2}}{1 + y^2} 
\end{equation}

and the second raw moment

\begin{equation}
\mathbb{E}(BF_{0,1}^2 \mid  \mathcal{M}_1) = \myint[_{-\infty}^\infty]{y}{ \frac{\left(\text{Cauchy}(y \mid  0, 1)\right)^2}{\text{Norm}(y \mid  0,1)}}
\end{equation}

which is infinite because

\begin{equation}
\lim_{y \to \pm\infty} \frac{\left(\text{Cauchy}(y \mid  0, 1)\right)^2}{\text{Norm}(y \mid  0,1)} = \lim_{y \to \pm\infty} \frac{\sqrt{2} \exp\left(\frac{y^2}{2}\right)}{\pi^\frac{3}{2}(1 + y^2)^2} = \infty 
\end{equation}

and hence also the variance is infinite.

In this case, \(\mathrm{Var}(BF_{0,1} \mid  \mathcal{M}_1)\) is
undefined. This is admittedly a bit extreme, as most models do not feature
Cauchy marginal data distributions. So let us assume that for a fixed,
known \(\mu\) we instead have
\(\mathcal{M}_0: y \sim \text{N}(\mu, 1)\).  Then the second raw moment is:

\begin{gather}    
\mathbb{E}(BF_{0,1}^2 \mid  \mathcal{M}_1) = \myint[_{-\infty}^\infty]{y}{ \frac{\left(\text{Norm}(y \mid  \mu, 1)\right)^2}{\text{Norm}(y \mid  0,1)}} = \myint[_{-\infty}^\infty]{y}{ \frac{\exp\left(\frac{x^2}{2} - (x - \mu)^2\right)}{\sqrt{2 \pi}}} \\
= \exp\left(\mu^2\right) \myint[_{-\infty}^\infty]{y}{\frac{\exp\left(-\frac{1}{2} (x - 2\mu)^2\right)}{\sqrt{2 \pi}}} = \exp\left(\mu^2\right)
\end{gather}

where in the last step, the integral is over the PDF of a normal
distribution with mean \(2\mu\) and standard deviation \(1\). Therefore
the variance of the Bayes factor is

\begin{equation}
\text{Var}(BF_{0,1}\mid \mathcal{M}_1) = \mathbb{E}(BF_{0,1}^2\mid \mathcal{M}_1) - E(BF_{0,1}\mid \mathcal{M}_1)^2 = \exp\left(\mu^2\right)  - 1.
\end{equation}

We note that this situation is completely symmetric in the sense that
\(\mathbb{E}(BF_{0,1}^2 \mid  \mathcal{M}_1) = \mathbb{E}(BF_{1,0}^2 \mid  \mathcal{M}_0)\)
so the problems are not specific to one ``direction'' of the Good check.

Now assume that \(\mu = 2\), then the standard error of the mean after
\(i\) iterations is \(\sqrt{\frac{\exp(4) - 1}{n}}\) so we expect to
need 5 360 simulations to keep the standard error below \(\frac{1}{10}\)
and 21 440 simulations to keep the standard error below \(\frac{1}{20}\)
and thus obtain estimates likely within \(\frac{1}{10}\) of the correct
value. This is
despite both marginal data distributions being well behaved and having
substantial overlap.

\begin{figure}

\centering{

\includegraphics[width=\textwidth]{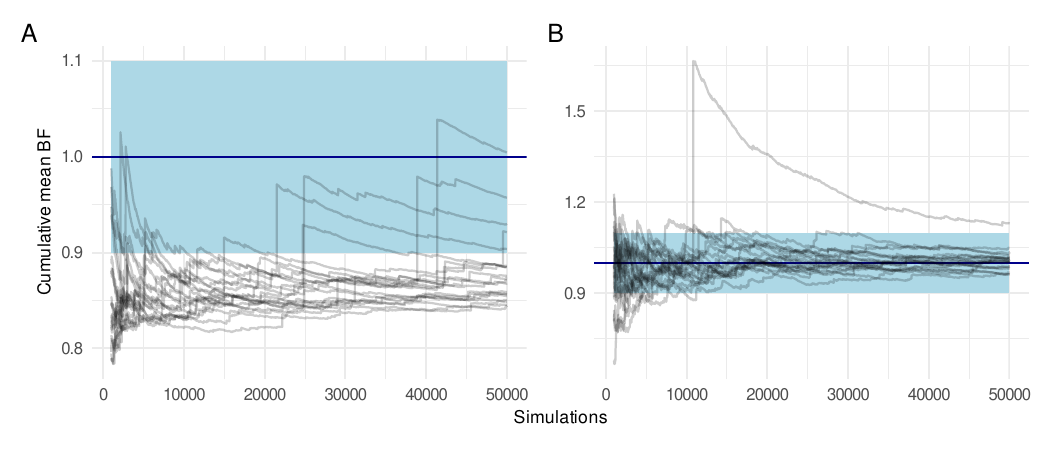}

}

\caption{\label{fig-good-convergence}Convergence of the good check when
a single datapoint is simulated from the standard normal and A)
\(\mathcal{M}_0: y \sim \text{Cauchy}(0, 1)\) --- here the variance is
infinite and we see no convergence at all or B)
\(\mathcal{M}_0: y \sim N(2, 1)\) where the average Bayes factor
eventually converges, but 50 000 simulations are not enough for reliable
convergence. Each line is the cumulative average from a single set of
simulations. The highlighted area shows average BF 0.9 - 1.1. Means from
first 1000 simulations are not shown due to their large variability.}

\end{figure}

Empirically, we can see the lack of any convergence for the Cauchy case
and very slow convergence for the normal case in
Figure~\ref{fig-good-convergence}. It follows, that the Good check
cannot reliably diagnose BF computation unless we know that the BF
distribution is well behaved.

\subsection*{SBC for the same models}\label{sbc-for-the-same-models}

\begin{figure}

\centering{

\includegraphics[width=\textwidth]{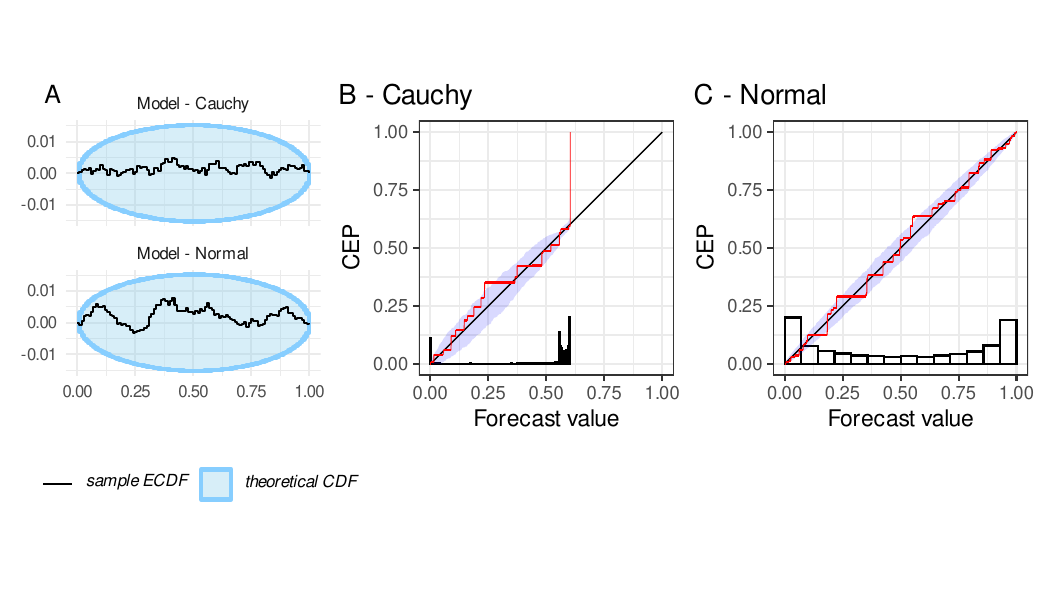}

}

\caption{\label{fig-good-sbc}ECDF difference plots from SBC (A) and
binary calibration plots (B, C) for Bayes factors when
\(\mathcal{M}_1: y \sim N(0, 1)\) and either
\(\mathcal{M}_0: y \sim \text{Cauchy}(0, 1)\) or
\(\mathcal{M}_0: y \sim N(2, 1)\). For the calibration plots, \calibrationplotinfo
}

\end{figure}%

On the contrary, SBC as well as prediction calibration and data-averaged
posterior have no problem and since the BF is computed correctly, show
that both scenarios pass the tests (to the precision available with the
10 000 simulations we ran). For SBC and prediction calibration see
Figure~\ref{fig-good-sbc}. Numerically for the Cauchy case we have 95\% CI for DAP difference from prior: -0.0032 -- 0.0038; miscalibration: 0.0008, 95\% quantile under null: 0.001. For the normal case we have 95\% CI for DAP difference from prior: -0.0141 -- 0.0005; miscalibration: 0.0011, 95\% quantile under null: 0.0013. 
In both cases SBC is sensitive to eCDF difference up to 0.015.

\section*{Appendix E - Additional empirical results}

\begin{table}[b!]
\small
\caption{\small Detailed results of validation checks where we found no problems in Bayes factor computation. DAP  95\% CI --- confidence interval for the difference between prior and data-averaged posterior, Miscalibration Value --- Observed value of miscalibration for the binary prediction calibration check as defined in \citet{dimitraidis2021_reliability}, Miscalibration Q95\% --- 95th percentile of bootstrapped distribution of miscalibration assuming perfect calibration. SBC sens. --- minimal absolute
difference between uniform CDF and empirical CDF SBC would detect. See Section 4 for details.}
\label{tab-successful-checks}
\centering
\begin{tabular}[t]{llrllll}
\toprule
Package & Scenario & \#Sims & DAP  95\% CI & \multicolumn{2}{c}{Miscalibration} & SBC\\
 & &  &  & Value & Q95\% & sens.\\
\midrule
bridgesampling & Turtles & 2000 & (-0.0382, 0.0145) & 0.0026 & 0.0037 & 0.036\\
 \ \ \ \ \  + Stan  & Turtles - post SBC & 2000 & (-0.0061, 0.0165) & 0.0034 & 0.0042 & 0.036\\
BayesFactor & Ranef presence & 10000 & (-0.0005, 0.013) & 0.001 & 0.0013 & 0.016\\
 & \texttt{ttestBF} & 50000 & (-0.0011, 0.0008) & 0.0003 & 0.0004 & 0.007\\
JAGS & Large model space & 1333 & (-0.0153, 0.0051) & 0.0021 & 0.0031 & 0.044\\
& \ \ \ (top model) & & & & & \\
\bottomrule
\end{tabular}
\end{table}

\begin{table}

\caption{\label{tab-power}Number of simulations required to obtain 80\% power to detect problems in BF computation across
the scenarios discussed in this paper. When large number of simulations was required, we calculated the statistics over steps of multiple simulations to keep computational demands manageable, hence the very round numbers in some rows. Bin. c. = binary calibration for the miscalibration metric, Brier = binary calibration with Brier score.}
\centering
\begin{tabular}[t]{llll}
\toprule
Scenario & Method & Variable & \# of sims (95\% CI)\\
\midrule
Poisson vs. NB, ignore data & SBC & log\_lik & 35 (32 -- 42)\\
Poisson vs. NB, ignore half & SBC & log\_lik & 240 (220 -- 270)\\
Poisson vs. NB,  & SBC & log\_lik & 180 (140 -- 230)\\
\ \ log BF noise with SD = 2 & SBC & model & 240 (190 -- 310)\\
 & Bin. c. & model & 60 (50 -- 70)\\
Poisson vs. NB, & SBC & log\_lik & 155 (145 -- 185)\\
 \ \ log BF bias = +2& SBC & model & 45 (40 -- 65)\\
 & Bin. c. & model & 15 (15 -- 20)\\
 & DAP & model & 10 (10 -- 10)\\
Turtles, normalization  & SBC & model & 274 (248 -- 350)\\
 \ \ constant divided by 2 & SBC & sigma & 192 (178 -- 270)\\
 & Bin. c. & model & 152 (100 -- 196)\\
 & DAP & model & 166 (126 -- 178)\\
Presence of random effects, & SBC & model & 250 (210 -- 350)\\
 \ \ constant prior & Bin. c. & model & 160 (150 -- 210)\\
 & DAP & model & 160 (130 -- 190)\\
JZS t-test, constant prior & SBC & model\_sd\_m1 & 1340 (1080 -- 1720)\\
Multiple models, trick & SBC & is\_model0 & 13600 (10800 -- 14800)\\
 & SBC & model & 15600 (12600 -- 17200)\\
 & Bin. c. & is\_model0 & 14400 (12800 -- 16600)\\
 & DAP & is\_model0 & 1565 (1305 -- 1795)\\
 & DAP & is\_model2 & 13800 (11000 -- 15000)\\
Multiple models, few draws & SBC & top\_model & 1280 (940 -- 1620)\\
 & Bin. c. & is\_model0 & 1680 (1480 -- 2160)\\
 & Bin. c. & top\_model & 1080 (860 -- 1320)\\
 & DAP & top\_model & 740 (640 -- 1040)\\
Large model space, & SBC & log\_lik & 3 (2 -- 3)\\
 \ \ shuffled & SBC & model\_index & 54 (46 -- 66)\\
 & Brier & m\_aggregated & 105 (70 -- 125)\\
 \bottomrule
\end{tabular}
\end{table}

\clearpage

\section*{\texorpdfstring{Appendix F - Miscalibration with surrogate
distributions for variance in
\texttt{ttestBF}}{Appendix F - Miscalibration with surrogate distributions for variance in ttestBF}}\label{miscalibration-with-surrogate-distributions-for-variance-in-ttestbf}

\begin{figure}[t!]
\centering{
\includegraphics[width=\figurescale\linewidth]{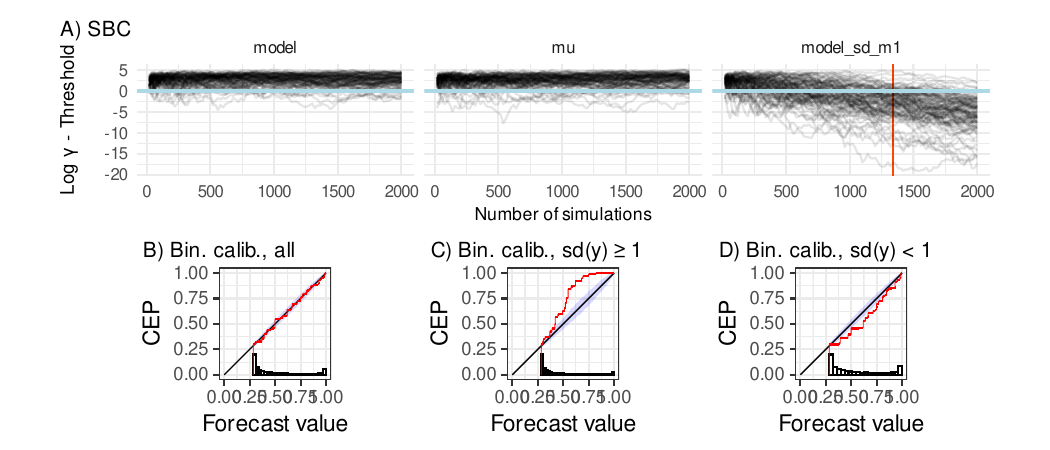}
}
\caption{\label{fig-ttest-fixed}\small Results of SBC when using a fixed value
of \(\sigma\) when simulating datasets for Bayesian t-test. A) Histories
of log-gamma statistics for the model index, posterior mean (mu) and
\(f(i, y) = (i - \frac{1}{2})(\text{sd}(y) - 1)\) (model\_sd\_m1), B) Calibration plot, C) and D) Separate calibration
plots for datasets with high/low standard deviation, 
\calibrationplotinfo}
\end{figure}

An example is the one-sample Bayesian t-test implemented in the
\texttt{ttestBF} function in the \texttt{BayesFactor} package. Here, the
\(\mathcal{M}_1\) model is:
\begin{equation}
\begin{aligned}
y_{1}, ..., y_n \mid \delta, \sigma &\sim N(\delta\sigma, \sigma), &
\delta \sim \text{Cauchy}\left(0, \frac{\sqrt{2}}{2}\right), & & \pi_\text{prior}\left(\sigma^2\right) = \frac{1}{\sigma^2}
\end{aligned}
\end{equation}
and the null model \(\mathcal{M}_0\) assumes \(\delta = 0\). Note the
improper Jeffrey's prior on variance. We will use \(n = 5\) in our
simulations to provide a reasonably wide distribution of observed BFs.
When we fix \(\sigma = 1\), the default statistics and calibration look
good. However, when we test SBC for a derived quantity that combines the
model index (\(i\)) with the observed standard deviation (SD) of the data, specifically
\(f(i, y) = (i - \frac{1}{2})(\text{SD}(y) - 1)\) we see a strong
apparent miscalibration (Figure~\ref{fig-ttest-fixed}A). Alternatively,
we can notice that for datasets where the observed SD is high, the
posterior probabilities in favor of the alternative hypothesis tend to
be too low, which is matched by a corresponding overestimation when the
observed SD is low (for datasets with \(\text{SD}(y) \geq 1\)
miscalibration p \textless{} 0.001; Figure~\ref{fig-ttest-fixed} B,C).
Similar, but more subtle miscalibration is also observed when we put a
proper prior on \(\sigma\) --- we tested
\(\pi_\text{prior}\left(\sigma^2\right) = \frac{1}{1 + \sigma^2}\) and
\(\sigma \sim \text{HalfCauchy}(0, 1)\), see Appendix E for detailed
results.

When we use posterior SBC with \(y_1 = (-1,1)\), the computed Bayes
factors pass all checks to high precision --- we have run 50 000
simulations and see no sign of problems including all derived quantities
(see Table~\ref{tab-successful-checks}). So we conclude that the \texttt{ttestBF} function in fact works well
and the above-mentioned violations of the SBC and binary prediction
calibration checks are simply a result of not matching the simulations
to the model.

%

\begin{figure}[t!]

\centering{

\includegraphics[width=\textwidth]{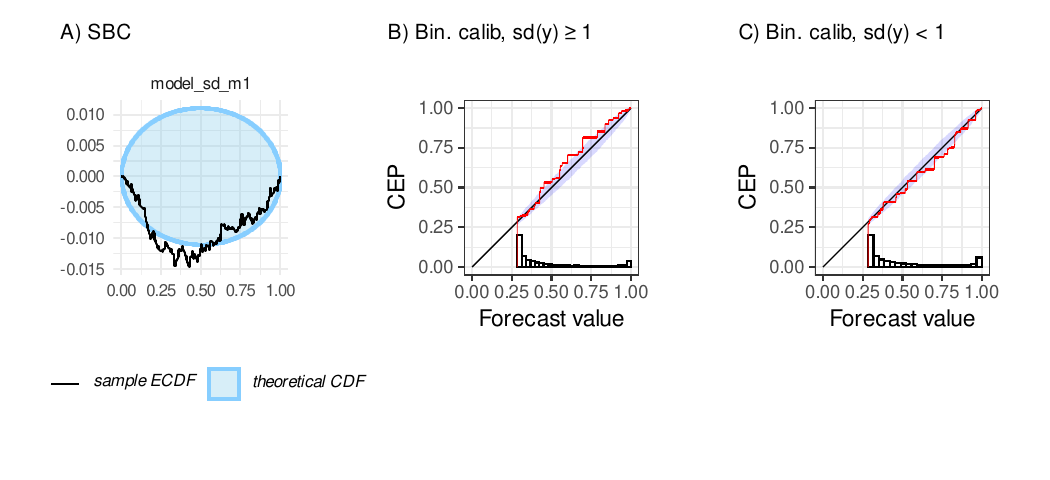}

}

\caption{\label{fig-ttest-inv1psquared}Results of SBC when using a
\(\pi_\text{prior}(\sigma^2) = \frac{1}{1 + \sigma^2}\) when simulating
datasets for Bayesian t-test. A) ECDF difference plot for the derived
quantity \(f(i, y) = (i - \frac{1}{2})(\text{sd}(y) - 1)\)
(model\_sd\_m1). B) and C) Separate binary prediction calibration plots
for datasets with high/low standard deviation. For the calibration plots, \calibrationplotinfo }

\end{figure}
\begin{figure}[b!]

\centering{

\includegraphics[width=\textwidth]{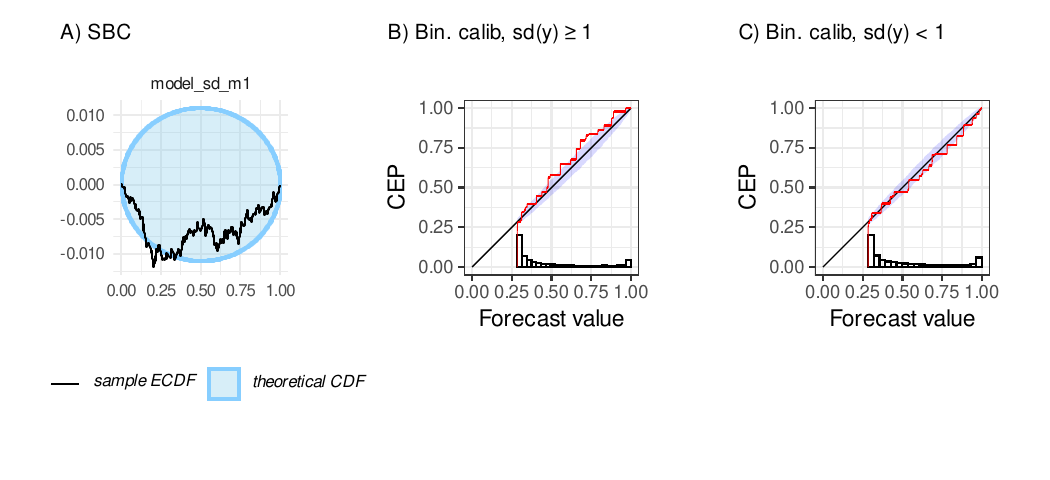}

}

\caption{\label{fig-ttest-cauchy}Results of SBC when using a
\(\sigma \sim \text{HalfCauchy}(0,1)\) when simulating datasets for
Bayesian t-test. A) ECDF difference plot for the derived quantity
\(f(i, y) = (i - \frac{1}{2})(\text{sd}(y) - 1)\) (model\_sd\_m1). B)
and C) Separate binary prediction calibration plots for datasets with
high/low standard deviation. For the calibration plots, \calibrationplotinfo}

\end{figure}

Recall the the improper prior for variance is
\(\pi_\text{prior}\left(\sigma^2\right) = \frac{1}{\sigma^2}\). First we
instead use a proper prior with similar form
\(\pi_\text{prior}(\sigma^2) = \frac{1}{1 + \sigma^2}\). We require 20
000 simulations to detect mild miscalibration which is visible in the
specifically designed derived quantity as well as when splitting the
results based on the standard deviation of the data, see
Figure~\ref{fig-ttest-inv1psquared}. We see similarly mild problems when
instead using \(\sigma \sim \text{HalfCauchy}(0,1)\), see
Figure~\ref{fig-ttest-cauchy}.

\clearpage

\section*{Appendix G - Comparison of Bayes Factors for the multiple models scenario}


\begin{figure}[b!]

\centering{

\includegraphics[width=\textwidth]{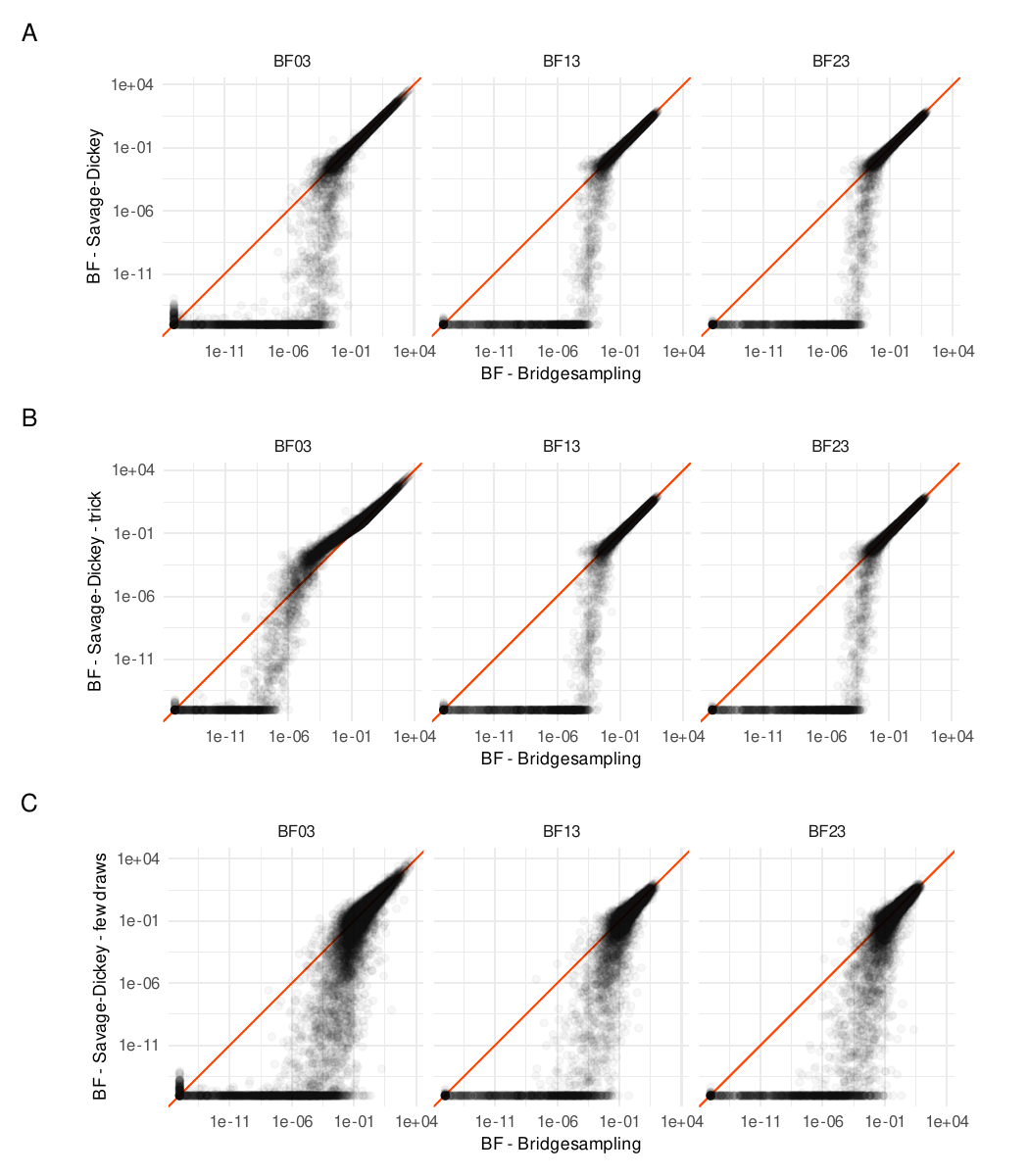}

}
\caption{\label{fig-multiple-bf-comparison} \small Bayes Factors against $\mathcal{M}_3$ in the 
multiple models scenario computed by bridge sampling compared to Savage - Dickey density ratio and the two problematic
variants thereof. Note that BF13 and BF23 are identical for the correct and trick scenarios. Each dot is a single simulation, orange diagonal line indicates perfect agreement.
}

\end{figure}%
\clearpage

\section*{Appendix H - High dimensional model space}

To show scalability to large model spaces, we employ the following model based on linear regression:

\begin{align}
    y_i \mid \beta, \sigma, m &\sim N\left(\sum_{j=1}^P{m_j \beta_j x_{j,i} }, \sigma\right) &  & & i &= 1, 2, \dots , N \notag \\
    \beta_j &\sim N(0, 1) &  m_j &\sim \text{Bernoulli}(u) & j &= 1,2,\dots, P \notag  \\
    u &\sim \text{Beta}(s_1, s_2) & \sigma &\sim \text{Exponential}(1)
\end{align}

This model encodes a binary decision to include each potential predictor (including the intercept) and represents a model space of the size $2^P$. For our demonstration we choose $N = 80$, $P = 100$, $s_1 = s_2 = 1$. We choose the predictor matrix $X$ such that it contains an intercept, marginal distribution of predictors beyond intercept is $N(0,1)$ and all pairwise predictor-to-predictor correlations are $0.8$. 
The model is fitted with JAGS using 4 chains with 5000 burnin iterations and then 400 000 sampling iterations, thinned by a factor of 100.

We have $P$ binary variables to check the correctness of the model selection. For large $P$, it becomes unreasonable to check all of them, so we instead have to aggregate the results in some way. One way is to use the top model indicator, though with large $P$, even the top model will often have small probability (quite frequently, each posterior draw corresponds to a different model), limiting its usefulness. To use a model index we need to choose an ordering of the models. Lexicographic ordering will give results that highly depend on the ordering of predictors (violations for predictors that are early and thus have high influence on the ordering will be easier to detect). We can somewhat reduce that dependence by choosing a graded lexicographic ordering, i.e. ordering first by the number of predictors and break ties lexicographically, though no general ``best'' ordering exists. Another implicit aggregation is by using the model log likelihood as a test quantity in SBC as that combines input from all predictor choices. Finally, we can compute p-values for individual miscalibration/data-averaged posterior checks for each $m_j$ and then aggregate them. The simplest is to take the smallest p-value after a family-wise error correction procedure like the Holm method.

Running 2000 simulations, 667 (33\%) of fits have signs of poor convergence and were removed. Among the converged fits, the posterior distribution appears calibrated to the precision available (e.g., for the top model indicator 95\% CI for DAP difference from prior: -0.015 -- 0.005; miscalibration: 0.0021, 95\% quantile under null: 0.0031; SBC sensitive to eCDF difference up to 0.044). To demonstrate detection of a global problem, we create a set of new posteriors by randomly shuffling posterior of 20 model indicator variables in each fit (choosing different indicators to shuffle for each fit). This is guaranteed to provide correct data-averaged posteriors for all model indicators as well as the top model indicator and we indeed see no violations there (data not shown).

The problem is the most quickly diagnosed by SBC for the log likelihood, requiring typically only 2 - 3 simulations.
The graded lexicographical ordering performs poorly in this setting, because its main part (the number of predictors included) is identical under the incorrect model, while the lexicographical ordering performs reasonably well as the incorrect model affects all predictors equally. We believe this is mostly an artifact of the chosen setup.

In this scenario the bootstrapped miscalibration test performs poorly and taking the smallest multiplicity corrected p-value over all model indicators attains only 35\% power to detect the problem after 500 simulations. Interestingly, its close alternative, the bootstrapped Brier score test also described in \cite{dimitraidis2021_reliability} performs quite well --- taking the smallest multiplicity corrected p-value discovers the problem in 105 simulations. In this test we perform the same bootstrap scheme (simulate binary predictions as if the observed probabilities were calibrated), but use those to build a null distribution for the Brier score. In all of our other examples, the Brier score test never performed better than the miscalibration test, so it is not reported elsewhere. We believe its good performance here is likely an artifact of the specific setup (shuffling model indices), we thus cannot recommend it in general. We note that using the bootstrapped tests with multiplicity correction requires generating a large number of bootstrap samples, to allow the individual tests to attain very low p-values. Otherwise the p-values will be guaranteed to be large after the correction.

\begin{figure}[t!]
\centering{

\includegraphics[width=\figurescale\linewidth]{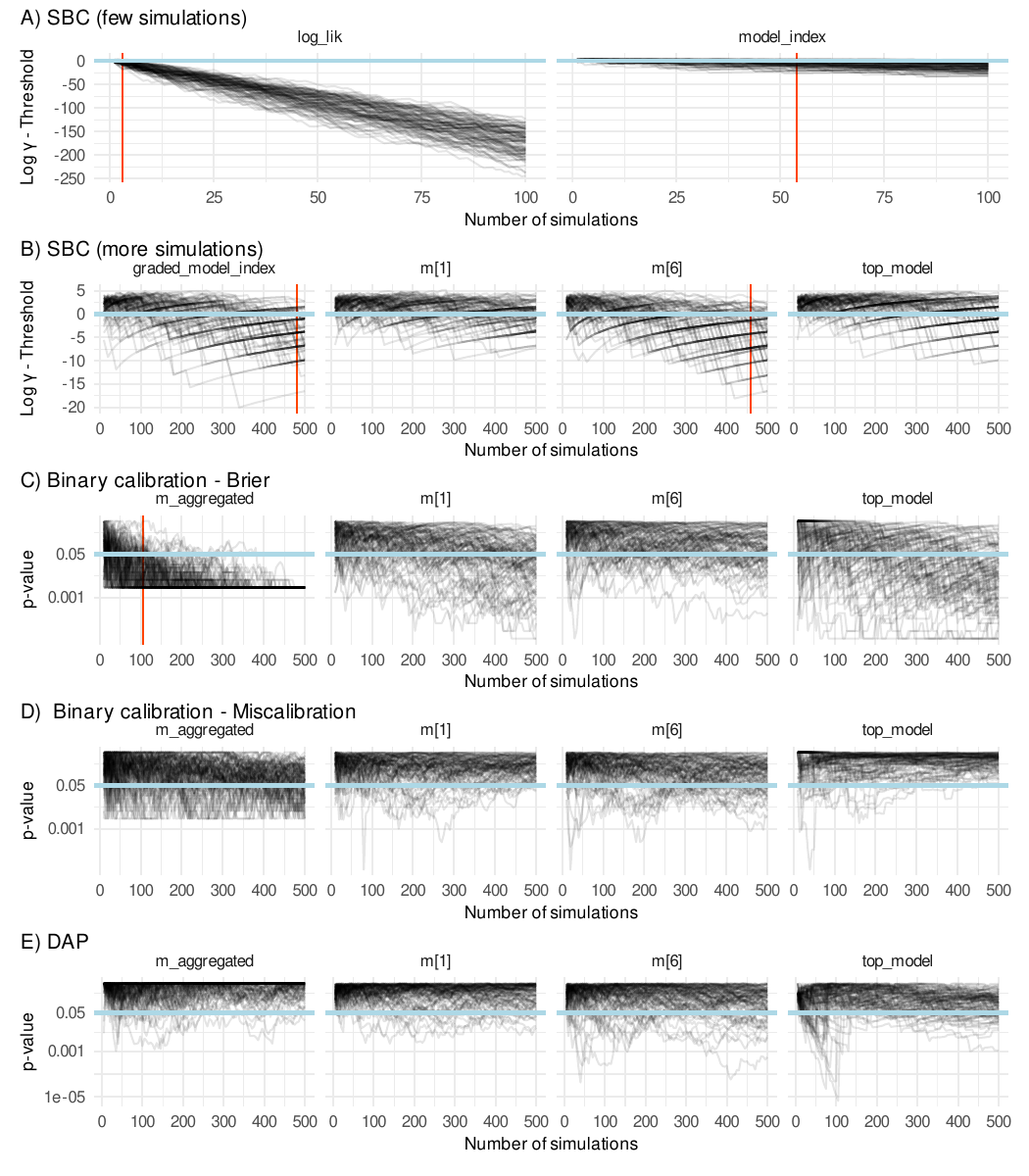}

}

\caption{\label{fig-large-model-space}\small Histories of check statistics for
the high-dimensional model space with shuffled model indices. A) and B) show the log-gamma statistic of
default SBC (note the different scales on both axes) - \texttt{log\_lik} is the log likelihood of the model, \texttt{model\_index} and \texttt{graded\_model\_index} is the (graded) lexicographical ordering of the models. \texttt{m[1]} and \texttt{m[6]} are two representative model indicators. C) and D) show the p-value of the bootstrapped Brier/miscalibration tests
of \citet{dimitraidis2021_reliability}~and E) is the p-value from t-test for
data-averaged posterior checking (DAP).  \historyinfo  
We see the problem detected quickly by SBC and Brier score test, while miscalibration test only very slowly gains power and DAP does not detect the problem at all.
}

\end{figure}

The data-averaged posterior check does not detect the problem at all. See Figure~\ref{fig-large-model-space} for the full histories.

\end{document}